\documentclass{elsart}

\usepackage{graphicx}

\usepackage{amssymb}

\newcommand{\eg}{{\it e.g.}}
\newcommand{\ie}{{\it i.e.}}

\newcommand{\beq}{\begin{equation}}
\newcommand{\eeq}{\end{equation}}
\newcommand{\bdm}{\begin{displaymath}}
\newcommand{\edm}{\end{displaymath}}
\newcommand{\E}{\mathcal{E}}
\newcommand{\Lim}[2]{\textrm{L}_{#1}\left({#2}\right)}
\newcommand{\mm}[2]{\textrm{minmod}\left({#1},{#2}\right)}
\newcommand{\sign}{\textrm{sign}}
\newcommand{\nf}{\mathcal{F}}

\begin{document}

\begin{frontmatter}



\title{An Unsplit Godunov Method for Ideal MHD via Constrained Transport in
       Three Dimensions}


\author[label1]{Thomas A. Gardiner}
\author[label1,label2]{\& James M. Stone}

\address[label1]{Department of Astrophysical Sciences \\ 
Princeton University \\
Princeton, NJ 08544}
\address[label2]{Program in Applied and Computational Mathematics \\
Princeton University \\
Princeton, NJ 08544}

\begin{abstract}
We present a single step, second-order accurate Godunov scheme for
ideal MHD which is an extension of the method described in \cite{GS05}
to three dimensions.  This algorithm combines the corner transport
upwind (CTU) method of Colella for multidimensional integration, and
the constrained transport (CT) algorithm for preserving the
divergence-free constraint on the magnetic field.  We describe the
calculation of the PPM interface states for 3D ideal MHD which must
include multidimensional ``MHD source terms'' and naturally respect
the balance implicit in these terms by the ${\bf\nabla\cdot B}=0$
condition.  We compare two different forms for the CTU integration
algorithm which require either 6- or 12-solutions of the Riemann
problem per cell per time-step, and present a detailed description of
the 6-solve algorithm.  Finally, we present solutions for test
problems to demonstrate the accuracy and robustness of the algorithm.
\end{abstract}

\begin{keyword}

\PACS 
\end{keyword}
\end{frontmatter}

\section{Introduction}
\label{sec:intro}

\par
In a previous paper \cite{GS05} we described a two-dimensional,
second-order accurate Godunov method for ideal MHD that evolves the
magnetic field using the Constrained Transport (CT)
\cite{Evans-Hawley} algorithm for preserving the divergence-free
constraint on the magnetic field.  In its simplest form, CT requires
area-averaged values of the magnetic field which are stored at cell
faces.  We argued that this is the most natural discrete
representation of the field in that the integral form of the induction
equation is based on area (rather than volume) averages, and therefore
the discrete form of the equations should respect this difference.
There are three important ingredients to our MHD algorithm: (1) a
modification of the piecewise parabolic method (PPM) reconstruction
step used to construct time-advanced estimates of the conserved
variables on cell faces that are fed to the Riemann solver to
incorporate multidimensional terms essential in MHD, (2) a new method
for constructing the fluxes (at cell edges) of the area-averaged
magnetic fields (at cell faces) from the fluxes returned by the
Riemann solver (at cell faces) of volume averaged magnetic fields (at
cell centers) which are based on the fundamental relationship between
the area- and volume-averaged variables, and (3) a directionally
unsplit integration algorithm based on the Corner Transport Upwind
(CTU) method \cite{Colella-CTU}.

\par
Through a series of test problems, we demonstrated the importance of
each of the ingredients to our algorithm.  In particular, we showed
through tests based on the advection of two-dimensional field loops
that our new methods for constructing the fluxes needed by the CT
algorithm are essential for stability, and are an improvement over
previous Godunov methods that use CT, e.g. \cite{Balsara-Spicer}.  We
also showed that by using the second-order accurate CTU integration
algorithm, a method could be constructed which has less numerical
dissipation and has the important property of reducing exactly to the
one-dimensional algorithm for plane-parallel, grid-aligned flows.
Since CT does not require costly solutions to elliptic equations, we
expect MHD Godunov schemes based on CT to be more cost effective that
those that use divergence-cleaning \cite{Crockett,Toth-divB}.  Given
the attractive properties of the method, it is of interest to extend
it to three-dimensions for use in applications.

\par
When directional splitting is used, the extension of Godunov methods
from two- to three-dimensions is usually trivial.  However,
directional splitting is unsuitable for MHD, because it is impossible
to enforce the divergence-free constraint between partial updates
unless all three components of the magnetic field are updated
together, which in turn violates the assumption basic to splitting
that each dimensional operator is independent and can be split from
the others.  As a result, in \cite{GS05} we adopted the unsplit CTU
integration scheme.  Even in hydrodynamics, the extension of CTU to
three-dimensions is not trivial \cite{Saltzman}.  For our MHD
algorithm, extension to three-dimensions requires modifying two of the
three ingredients of the method, in particular (1) the PPM
reconstruction algorithm must be modified to include multidimensional
terms for MHD in such a way as to respect a \emph{balance} law implied
by the ${\bf\nabla\cdot B}=0$ condition, and (2) the CTU algorithm
must be modified to include source terms as well as the transverse
flux gradient terms.  The primary purpose of this paper is to describe
in detail these modifications and to demonstrate that the resulting
algorithm is both accurate and robust.

\par
We extend our MHD test suite to three-dimensions to demonstrate the
accuracy and fidelity of our method.  We find that, once again, the
passive advection of a multidimensional field loop is a challenging
test of finite volume methods for MHD.  In particular, for a field
loop confined to the $(x,y)$-plane in three-dimensions advected with a
constant velocity with $v_z \ne 0$, the vertical component of the
magnetic field $B_z$ will evolve unless care is made to ensure the
multidimensional balance of MHD source terms in both the PPM
characteristic tracing step and the transverse flux gradient update
step.  In fact, this observation leads to a useful definition of the
appropriate difference stencil on which the divergence-free constraint
must be maintained.  If $\nabla \cdot {\bf B} = 0$ on a stencil which
is different from that used to construct the fluxes of $B_z$, the
latter will show unphysical evolution in this test for conservative
algorithms.  On the other hand, if a numerical method keeps $B_z$
constant to round-off error on the test, it must preserve the
divergence-free constraint on the appropriate stencil. Moreover, this
test is another demonstration that it is {\em essential} to maintain
the divergence-free constraint {\em exactly} in MHD, as was originally
emphasized by \cite{Brackbill-Barnes}.  This test, along with linear
wave convergence tests, a test based on the propagation and
convergence of nonlinear, circularly polarized Alfv\'en waves, and
multidimensional blast wave tests are all presented in section 6.

\par
The paper is organized as follows.  In section 2, we write down the
equations of ideal MHD solved by our method, and describe the
finite-volume discretization of mass, momentum, and energy, and the
finite-area discretization of the magnetic field.  In section 3, we
describe our extension of the PPM reconstruction algorithm to
three-dimensional MHD.  In section 4, we briefly review the upwind CT
algorithms introduced in \cite{GS05}, while in section 5, we describe
two formulations for the CTU integration algorithm to 3D.  In section
6 we present the results of our test suite, while in section 7 we
conclude.

\section{Ideal Magnetohydrodynamics and Constrained Transport}
\label{sec:MHD-CT}

The equations of ideal magnetohydrodynamics (MHD) can be written in
conservative form as
\begin{eqnarray}
\frac{\partial \rho}{\partial t} + 
{\bf\nabla\cdot} \left(\rho{\bf v}\right) & = & 0 
\label{eq:cons_mass} \\
\frac{\partial \rho {\bf v}}{\partial t} + 
{\bf\nabla\cdot} \left(\rho{\bf vv} - {\bf BB}\right) +
{\bf \nabla} P^* & = & 0 \\
\frac{\partial {\bf B}}{\partial t} + 
{\bf \nabla \times} ({\bf B \times v}) & = & 0 \\
\label{eq:induct}
%
%
\frac{\partial E}{\partial t} + 
\nabla\cdot((E + P^*) {\bf v} - {\bf B} ({\bf B \cdot v})) & = & 0 
\label{eq:cons_energy}
\end{eqnarray}
where $\rho$ is the mass density, $\rho{\bf v}$ the momentum density,
${\bf B}$ the magnetic field, and $E$ the total energy density.  The
total pressure $P^* \equiv P + ({\bf B \cdot B})/2$ where $P$ is the
gas pressure.  This system of equations is closed with the addition of
an equation of state which relates the pressure and density to the
internal energy,
\beq
\epsilon \equiv E - \rho({\bf v \cdot v})/2 - ({\bf B \cdot B})/2 ~.
\eeq
Throughout this paper we will assume an ideal gas equation of state
for which $P = (\gamma - 1) \epsilon$, where $\gamma$ is the ratio of
specific heats.  Note that we have chosen a system of units in which
the magnetic permeability $\mu=1$.

\par
In addition to the evolutionary conservation laws, equations
(\ref{eq:cons_mass}) through (\ref{eq:cons_energy}), the magnetic
field must also obey the divergence free constraint,
i.e. ${\bf\nabla\cdot B}=0$.  Here, as in \cite{GS05}, this is
accomplished using the method of constrained transport (CT). In this
method one starts from the differential form of the induction equation
(\ref{eq:induct}) and constructs an integral relation by area
averaging the components of ${\bf B}$ normal to the grid cell faces
over the respective face and applying Stoke's theorem.  Analogous to
the finite volume method, the resulting integral relation forms the
basis of the numerical evolutionary equation.  Note that one immediate
consequence of CT is that the ${\bf\nabla\cdot B}=0$ constraint on the
magnetic field is satisfied in an integral sense over the smallest
discretization scale, the grid cell.  The volume averaged magnetic
field components, which for example are necessary to calculate the
internal energy, are defined equal to the average of the interface
averaged components.

\par
In this paper we will assume a regular, three dimensional, Cartesian
grid.  We will use the standard notation that grid cell $(i,j,k)$ is
centered at $(x_i,y_j,z_k)$ and has a size $(\delta x,\delta y,\delta
z)$.  Time levels will be denoted by a superscript and interface
values will be denoted by half increments to the index, \eg{} the
volume averaged $x$-component of the magnetic field at time $t^n$ is
defined to be
\beq
B_{x,i,j,k}^n \equiv
\frac{1}{2} \left( B_{x,i-1/2,j,k}^n + B_{x,i+1/2,j,k}^n \right) ~.
\eeq
%

\section{Calculating the Interface States}
\label{sec:lr_states}

\par
In this section we describe the calculation of the ``interface
states'' in the PPM algorithm for ideal MHD in three dimensions.  The
PPM interface state algorithm is based upon the idea of dimensional
splitting, and as a result is a one-dimensional algorithm including
both spatial reconstruction and a characteristic evolution of the
linearized system in primitive variables.  For ideal MHD, however, it
was shown in {}\cite{GS05} that it is necessary to include
multidimensional terms when calculating the interface states.  The
three dimensional interface state algorithm is thus a generalization
of the two-dimensional algorithm which for consistency must reduce to
the two- and one-dimensional algorithm in the appropriate limits.  The
interface states in the PPM algorithm are typically calculated by
evolving the system of equations in primitive variables.  Consider the
induction equation, which in component form is
\beq
\frac{\partial B_x}{\partial t} 
+ \frac{\partial}{\partial y} \left(v_y B_x - B_y v_x \right)
+ \frac{\partial}{\partial z} \left(v_z B_x - B_z v_x \right) = 0
\label{eq:Bx_induction}
\eeq
\beq
\frac{\partial B_y}{\partial t} 
+ \frac{\partial}{\partial x} \left(v_x B_y - B_x v_y \right) 
+ \frac{\partial}{\partial z} \left(v_z B_y - B_z v_y \right) = 0
\label{eq:By_induction}
\eeq
\beq
\frac{\partial B_z}{\partial t} 
+ \frac{\partial}{\partial x} \left(v_x B_z - B_x v_z \right) 
+ \frac{\partial}{\partial y} \left(v_y B_z - B_y v_z \right) = 0 ~.
\label{eq:Bz_induction}
\eeq
In these equations there are terms proportional to $\partial
B_x/\partial x$, $\partial B_y / \partial y$ and $\partial B_z
/\partial z$ which we will refer to as ``MHD source terms".  (When the
system of equations for MHD is written in primitive variables, these
source terms only appear in the induction equation.  As a result we
will not discuss the remaining MHD equations in this section.)  The
question before us is: which terms in the induction equation need to
be included in the calculation of the interface states?  In what
follows we specialize to the calculation of the $x$-interface states;
the $y$- and $z$-interface state calculation follows by symmetry.

\subsection{2D MHD Interface State Algorithm}
\label{sec:2D_lr_states}

\par
Before constructing the three-dimensional interface algorithm, it is
instructive to recall the two-dimensional algorithm presented in
{}\cite{GS05}.  For the two-dimensional $(x,y)$-case, the induction
equation for $B_z$ was simplified using the ${\bf \nabla \cdot B}=0$
condition eliminating the MHD source terms from the evolutionary
equation for $B_z$.  The result is the following set of equations for
calculating the $x$-interface states in 2D

\beq
\frac{\partial B_x}{\partial t} = 0
\label{eq:Bx_induction_lr_2D}
\eeq
\beq
\frac{\partial B_y}{\partial t} 
+ \frac{\partial}{\partial x} \left(v_x B_y - B_x v_y \right) = 0
\label{eq:By_induction_lr_2D}
\eeq
\beq
\frac{\partial B_z}{\partial t} 
+ \frac{\partial}{\partial x} \left(v_x B_z \right) 
- B_x \frac{\partial v_z}{\partial x} = 0 ~.
\label{eq:Bz_induction_lr_2D}
\eeq
Utilizing the ${\bf \nabla \cdot B}=0$ condition to eliminate the
source terms from the evolutionary equation for $B_z$ bestows a very
important property on the calculation of the interface states, namely
the balance of the MHD source terms is exactly and explicitly
included.  The importance of this step can be easily understood by
considering the advection of a magnetic field loop initially confined
to the $(x,y)$-plane (\ie{} $B_z=0$) with ${\bf v}=$ constant, $v_z
\neq 0$, and $\beta = 2 P/B^2 \gg 1$.  If the MHD source terms had not
been eliminated from equation (\ref{eq:Bz_induction}) prior to
dimensional splitting and the following equation
\beq
\frac{\partial B_z}{\partial t} 
+ \frac{\partial}{\partial x} \left(v_x B_z - B_x v_z \right) = 0
\label{eq:Bz_induction_lr_2D_bad}
\eeq
was used instead of equation (\ref{eq:Bz_induction_lr_2D}) the
$x$-interface state would include an erroneous $B_z$ evolution owing
to the term $v_z \left(\partial B_x/\partial x\right)$.  Experience
shows that this error is not eliminated when updating the interface
states due to transverse flux gradients in the CTU algorithm, leading
to steady, secular growth of $B_z$ which effectively warps the field
loop.  The important point to note here is that the \emph{balance} of
the MHD source terms resulting from the ${\bf \nabla \cdot B}=0$
condition must be accurately represented in the calculation of the
interface states.

\subsection{3D MHD Interface State Algorithm}
\label{sec:3D_lr_states}

\par
The 3D interface algorithm we construct is designed to explicitly
incorporate the potential balance between the MHD source terms and to
reduce exactly to the 2D interface states algorithm in the limit that
the problems is two-dimensional and grid aligned.  The essential idea
is to rewrite the induction equation as follows prior to applying the
idea of directional splitting.
\begin{eqnarray}
\frac{\partial B_x}{\partial t} 
& + & \left\{ \frac{\partial}{\partial y} \left(v_y B_x - B_y v_x \right)
- v_x \Lim{xy}{\frac{\partial B_z}{\partial z}}
\right\} \nonumber \\
& + & \left\{ \frac{\partial}{\partial z} \left(v_z B_x - B_z v_x \right)
- v_x \Lim{xz}{\frac{\partial B_y}{\partial y}}
\right\} = 0
\label{eq:Bx_induction_lim}
\end{eqnarray}
\begin{eqnarray}
\frac{\partial B_y}{\partial t} 
& + & \left\{ \frac{\partial}{\partial x} \left(v_x B_y - B_x v_y \right) 
- v_y \Lim{yx}{\frac{\partial B_z}{\partial z}} 
\right\} \nonumber \\
& + & \left\{ \frac{\partial}{\partial z} \left(v_z B_y - B_z v_y \right) 
- v_y \Lim{yz}{\frac{\partial B_x}{\partial x}}
\right\} = 0
\label{eq:By_induction_lim}
\end{eqnarray}
\begin{eqnarray}
\frac{\partial B_z}{\partial t} 
& + & \left\{ \frac{\partial}{\partial x} \left(v_x B_z - B_x v_z \right) 
- v_z \Lim{zx}{\frac{\partial B_y}{\partial y}}
\right\} \nonumber \\
& + & \left\{ \frac{\partial}{\partial y} \left(v_y B_z - B_y v_z \right)
- v_z \Lim{zy}{\frac{\partial B_x}{\partial x}}
\right\} = 0 ~,
\label{eq:Bz_induction_lim}
\end{eqnarray}
where we've added a limited amount of the transverse MHD source term
to each component of the electric field gradient and grouped terms
according to the fashion in which they will be split.  The
mathematical form of the limiter functions, e.g. $\textrm{L}_{xy}$, is
determined by imposing constraints on the directionally split and
unsplit system.  Clearly, to recover the induction equation we have
\beq
\Lim{xy}{\frac{\partial B_z}{\partial z}} = -
\Lim{xz}{\frac{\partial B_y}{\partial y}} 
\label{eq:lim_symmetry}
\eeq
et cetra.  Directionally split, we obtain the following system for the
$x$-coordinate direction
\beq
\frac{\partial B_x}{\partial t} = 0 
\label{eq:Bx_induction_lim_s}
\eeq
\beq
\frac{\partial B_y}{\partial t} 
+ \frac{\partial}{\partial x} \left(v_x B_y - B_x v_y \right) 
- v_y \Lim{yx}{\frac{\partial B_z}{\partial z}} = 0
\label{eq:By_induction_lim_s}
\eeq
\beq
\frac{\partial B_z}{\partial t} 
+ \frac{\partial}{\partial x} \left(v_x B_z - B_x v_z \right) 
- v_z \Lim{zx}{\frac{\partial B_y}{\partial y}} = 0 ~.
\label{eq:Bz_induction_lim_s}
\eeq
To determine the form of the limiter functions, we minimize the
magnitude of the sum of the MHD source terms.  For equation
(\ref{eq:Bz_induction_lim_s}) we find
\beq
\Lim{zx}{\frac{\partial B_y}{\partial y}} = 
\mm{-\frac{\partial B_x}{\partial x}}{\frac{\partial B_y}{\partial y}} 
\label{eq:Bz_x_lim}
\eeq
where the $\textrm{minmod}$ function is defined as
\beq
\mm{x}{y} = \left \{
\begin{array}{ll}
\sign(x) \min(|x|,|y|) & \textrm{if }xy > 0 \\
0 & \textrm{otherwise.}
\end{array}
\right .
\eeq
Note that this limiter function satisfies the constraint identified in
equation (\ref{eq:lim_symmetry}).  The mathematical form of the
remaining limiter functions in equations (\ref{eq:Bx_induction_lim} -
\ref{eq:Bz_induction_lim}) is given by cyclic permutation of $(x,y,z)$
in equation (\ref{eq:Bz_x_lim}) and application of the constraint
noted in equation (\ref{eq:lim_symmetry}).

\par
There is also a simple physical argument for why the limiter function
takes the form described by equation (\ref{eq:Bz_x_lim}).  Considering
equation (\ref{eq:Bz_induction}), if $\left(\partial B_x / \partial
x\right)$ and $\left(\partial B_y / \partial y\right)$ have opposite
signs, but not necessarily the same magnitude we wish to incorporate
the balance of these two MHD source terms by adding and subtracting
the term with the smaller magnitude so that the resulting (reduced)
MHD source term is associated with only one of the flux gradients.
If, on the other hand, these derivatives have the same sign, then
there is no balance between the source terms and the induction
equation should be unmodified.  This is precisely the result of the
$\textrm{minmod}$ limited source term in equation (\ref{eq:Bz_x_lim}).

\par
Finally, we note that using the properties of the $\textrm{minmod}$
function and the ${\bf\nabla\cdot B}=0$ condition, equations
(\ref{eq:Bx_induction_lim_s}-\ref{eq:Bz_induction_lim_s}) can be
simplified to
\beq
\frac{\partial B_x}{\partial t} = 0
\label{eq:Bx_induction_lr_3D_2}
\eeq
\beq
\frac{\partial B_y}{\partial t} 
+ \frac{\partial}{\partial x} \left( v_x B_y \right) 
- B_x \frac{\partial v_y}{\partial x} 
- v_y \mm{\frac{\partial B_x}{\partial x}}{-\frac{\partial B_y}{\partial y}}
= 0
\label{eq:By_induction_lr_3D_2}
\eeq
\beq
\frac{\partial B_z}{\partial t} 
+ \frac{\partial}{\partial x} \left( v_x B_z \right)
- B_x \frac{\partial v_z}{\partial x} 
- v_z \mm{\frac{\partial B_x}{\partial x}}{-\frac{\partial B_z}{\partial z}}
= 0
\label{eq:Bz_induction_lr_3D_2}
\eeq
for calculating the $x$-interface states.  As a practical matter,
these limited MHD source terms are evaluated in terms of the cell
average of the magnetic field gradients, i.e. for equation
(\ref{eq:Bz_induction_lr_3D_2}) in cell $(i,j,k)$ we use
\beq
\mm{\frac{B_{x,i+1/2,j,k} - B_{x,i-1/2,j,k}}{\delta x}}
{\frac{B_{z,i,j,k-1/2} - B_{x,i,j,k+1/2}}{\delta z}}
\eeq
The equations for the $y$- and $z$-interface states follow from cyclic
permutations of $(x,y,z)$.  In the limiting two dimensional case of
either $\partial/\partial y=0$ or $\partial/\partial z=0$ this
approach reduces to the interface state algorithm outlined in
\S\ref{sec:2D_lr_states}.  Moreover, in the limiting two dimensional
case of $\partial/\partial x=0$, the $x$-interface state will equal
the cell center state, just what one expects from one and two
dimensional calculations.  As a result, this algorithm for calculating
the interface states preserves the $B_z=0$ condition for the {\em
gedanken} experiment described in
\S\ref{sec:2D_lr_states} involving the advection of a magnetic field
loop.

\section{Constrained Transport (CT) Algorithm}
\label{sec:CT_Alg}

\par
The coupling of a Godunov, finite volume algorithm with the method of
CT requires an algorithm for constructing the grid cell edge averaged
electric fields (emfs) from the Godunov fluxes.  This algorithm is
typically referred to as a CT algorithm.  The process of applying a CT
algorithm to calculate the CT emfs from the Godunov emf's can be
described as a predictor / corrector process where the Godunov emf's
are the predictor values and the resulting CT emfs are the corrector
values.  In {}\cite{GS05} a simple framework for constructing CT
algorithms was presented and a few CT algorithms were constructed and
tested with the $\E^c$ CT algorithm selected as having the best
properties.  The $\E^c$ CT algorithm is constructed in such a way as
contain an upwind bias (according to the contact mode) and to reduce
to the correct Godunov emf for grid-aligned, plane-parallel flows.  In
this paper we will also use the $\E^c$ CT algorithm which for the sake
of completeness we briefly review here.

\par
Consider for the moment the calculation of the $z$-component of the
electric field at the grid cell edge $(i+1/2, j+1/2, k)$. (The
calculation of the $x$- and $y$-components of the CT electric fields
follows an analogous procedure.)  The CT algorithms described in
{}\cite{GS05} compute the CT electric field at this location from the
four neighboring face center electric field components (Godunov
fluxes) as well as estimates of the gradients of the electric field as
follows
\begin{eqnarray}
\E_{z,i+1/2,j+1/2,k} & = &
\frac{1}{4} \left(
\E_{z,i+1/2,j,k} + \E_{z,i+1/2,j+1,k} + \E_{z,i,j+1/2,k} + \E_{z,i+1,j+1/2,k} 
\right) \nonumber \\
& + & \frac{\delta y}{8} \left(
\left(\frac{\partial \E_z}{\partial y}\right)_{i+1/2,j+1/4,k} - 
\left(\frac{\partial \E_z}{\partial y}\right)_{i+1/2,j+3/4,k}
\right) \nonumber \\
& + & \frac{\delta x}{8} \left(
\left(\frac{\partial \E_z}{\partial x}\right)_{i+1/4,j+1/2,k} - 
\left(\frac{\partial \E_z}{\partial x}\right)_{i+3/4,j+1/2,k}
\right) ~.
\label{eq:2d_CT_Ez}
\end{eqnarray}
To complete this CT algorithm we need to specify a way to calculate
the derivatives of $\E_z$ on the grid cell face.  The $\E^c$ CT
algorithm computes the electric field gradient at the grid cell face
by selecting the ``upwind'' direction according to the contact mode,
i.e.
\beq
\left(\frac{\partial \E_z}{\partial y}\right)_{i+1/2,j+1/4,k} = \left \{
\begin{array}{ll}
(\partial \E_z/\partial y)_{i,j+1/4,k} & \textrm{for}~ v_{x,i+1/2,j,k} > 0 \\
(\partial \E_z/\partial y)_{i+1,j+1/4,k} & \textrm{for}~ v_{x,i+1/2,j,k} < 0 \\
\frac{1}{2}\left(
(\partial \E_z/\partial y)_{i,j+1/4,k} + (\partial \E_z/\partial y)_{i+1,j+1/4,k}
\right) & \textrm{otherwise} 
\end{array} 
\right .
\label{eq:Ndup_CT_Ez}
\eeq
with an analogous expression for the $(\partial \E_z/\partial x)$.
The final detail involves the definition of the electric field
derivatives in equation (\ref{eq:Ndup_CT_Ez}).  These are computed
using the face centered electric fields (Godunov fluxes) and a cell
center ``reference'' value $\E^r_{z,i,j,k}$, e.g.
\beq
\left(\frac{\partial \E_z}{\partial y}\right)_{i,j+1/4,k} = 
2\left(\frac{\E_{z,i,j+1/2,k} - \E^r_{z,i,j,k}}{\delta y}\right) ~.
\label{eq:CT_Ez_der}
\eeq
In the 2D MHD CTU algorithm described in {}\cite{GS05} and the 3D
version described here, the cell center reference electric field
$\E^r_{z,i,j,k}$ is computed using the cell center state at an
appropriate time level.  For the first interface flux calculation,
using the interface states described in {}\S\ref{sec:3D_lr_states},
the cell center reference electric field is computed using the cell
center state $q_{i,j,k}^n$ when integrating from time $t^n$ to
$t^{n+1}$.  The calculation of the cell center reference electric
field in subsequent steps of the integration algorithm use time
advanced states and will be described later in connection with the
integration algorithm.

\section{Corner Transport Upwind Algorithm}
\label{sec:CTU_alg}

In this section we are interested in applying the CTU algorithm to the
system of equations for ideal MHD.  The CTU algorithm was originally
described by Colella \cite{Colella-CTU} as an unsplit, 2D finite
volume algorithm for solving hyperbolic systems of conservation laws.
The 3D generalization of the CTU algorithm was subsequently presented
by Saltzman {}\cite{Saltzman}.  The CTU algorithm is generally set
within a predictor-corrector formalism and utilizes PPM
{}\cite{Colella-Woodward-PPM} when applied to Euler's equations - the
archetypical system.

\par
Prior to delving into the details of the numerical algorithms, it is
worth while pointing out that the system of equations for ideal MHD
differs from Euler's equations in non-trivial ways which are very
important when applying the CTU algorithm to MHD.  First, a straight
forward application of the directional splitting technique to the MHD
equations in primitive and conservative variables results in an
incompatible set of equations.  This results in the need to
incorporate source terms in the transverse flux gradient corrections
in the 2D and 3D CTU integration algorithm {}\cite{GS05} since the PPM
interface state algorithm uses the primitive variable form of the
equations.  Second, the treatment of the multidimensional MHD source
terms in the MHD PPM interface states algorithm, described in
{}\S\ref{sec:lr_states}, results in the need to incorporate source
terms in the transverse flux gradient updates to the transverse
components of the magnetic field at the interfaces.  These details
follow from the balance between multidimensional flux gradients
imposed by the ${\bf\nabla\cdot B}=0$ condition.

\par
In this section we present two variants of the CTU integration
algorithm.  In \S\ref{sec:12_solve} we present a brief, functional
description of the 3D CTU algorithm as described by Saltzman
{}\cite{Saltzman} which we refer to here as the 12-solve algorithm
since it requires 12 solutions to the Riemann problem per zone per
time step.  We will discuss the challenges associated with adapting
this algorithm to the equations of ideal MHD.  However, for a variety
of reasons, the principle one being the complexity of the algorithm,
we will not present the algorithmic elements for the 12-solve MHD CTU
algorithm in detail.  In {}\S\ref{sec:6_solve} we present a simple
variant on the CTU algorithm which requires only 6 solutions to the
Riemann problem per zone per time step and describe this 6-solve
algorithm in detail.  We summarize with a discussion of the strengths
and weaknesses of this algorithm relative to the 12-solve CTU
algorithm as a prelude to \S\ref{sec:tests} where we present a variety
of results comparing the 6-solve and 12-solve MHD CTU algorithms.

\subsection{12-solve CTU}
\label{sec:12_solve}

In this subsection we present a functional description of the 12-solve
CTU algorithm as constructed for Euler's equations.  This serves the
goal of making the discussion more self-contained as well as allowing
us to directly point out where particular elements of the integration
algorithm pose challenges when applied to ideal MHD.  For a more
detailed description of the algorithm, or the theoretical
underpinnings, see {}\cite{Colella-CTU,Miller-Colella,Saltzman}.

\par
We begin by choosing a numerical flux function $\nf(q_L, q_R)$ which
is assumed to return a suitably accurate solution for the flux
obtained by solving the Riemann problem associated with $q_L$ and
$q_R$, the left and right states.  The 12-solve CTU algorithm can then
be described as follows.

\par
Step 1, calculate the left and right PPM interface states
$q_{Lx,i+1/2,j,k}^*$, $q_{Rx,i+1/2,j,k}^*$, $q_{Ly,i,j+1/2,k}^*$,
$q_{Ry,i,j+1/2,k}^*$, $q_{Lz,i,j,k+1/2}^*$ and $q_{Rz,i,j,k+1/2}^*$
and the associated interface fluxes
\begin{eqnarray}
F_{x,i+1/2,j,k}^* &=& \nf_x(q_{Lx,i+1/2,j,k}^*,q_{Rx,i+1/2,j,k}^*) \\ 
F_{y,i,j+1/2,k}^* &=& \nf_y(q_{Ly,i,j+1/2,k}^*,q_{Ry,i,j+1/2,k}^*) \\ 
F_{z,i,j,k+1/2}^* &=& \nf_z(q_{Lz,i,j,k+1/2}^*,q_{Rz,i,j,k+1/2}^*) .
\end{eqnarray}

\par
Step 2, for each interface state calculate two interface states
evolved by $\delta t/3$ of a single transverse flux gradient, i.e.
\begin{eqnarray}
q_{Lx,i+1/2,j,k}^{*|y} &=& q_{Lx,i+1/2,j,k}^* + \frac{\delta t}{3 \delta y}
\left(F_{y,i,j-1/2,k}^* - F_{y,i,j+1/2,k}^* \right) \\
q_{Rx,i+1/2,j,k}^{*|y} &=& q_{Rx,i+1/2,j,k}^* + \frac{\delta t}{3 \delta y}
\left(F_{y,i+1,j-1/2,k}^* - F_{y,i+1,j+1/2,k}^* \right) \\
q_{Lx,i+1/2,j,k}^{*|z} &=& q_{Lx,i+1/2,j,k}^* + \frac{\delta t}{3 \delta z}
\left(F_{z,i,j,k-1/2}^* - F_{z,i,j,k+1/2}^* \right) \\
q_{Rx,i+1/2,j,k}^{*|z} &=& q_{Rx,i+1/2,j,k}^* + \frac{\delta t}{3\delta z}
\left(F_{z,i+1,j,k-1/2}^* - F_{z,i+1,j,k+1/2}^* \right) 
\end{eqnarray}
with $y$- and $z$-interface states being defined in an equivalent
manner by cyclic permutation of $(x,y,z)$ and $(i,j,k)$.  For each of
the $\delta t/3$ updated interface states, calculate the associated
flux, giving the two $x$-interface fluxes
\begin{eqnarray}
F_{x,i+1/2,j,k}^{*|y} &=& \nf_x(q_{Lx,i+1/2,j,k}^{*|y},q_{Rx,i+1/2,j,k}^{*|y})\\
F_{x,i+1/2,j,k}^{*|z} &=& \nf_x(q_{Lx,i+1/2,j,k}^{*|z},q_{Rx,i+1/2,j,k}^{*|z})
%
%
\end{eqnarray}
and similar expressions for the $y$- and $z$-interface fluxes.

\par
Step 3, at each interface evolve the PPM interface states by $\delta
t/2$ of the transverse flux gradients, i.e.
\begin{eqnarray}
q_{Lx,i+1/2,j,k}^{n+1/2} &=& q_{Lx,i+1/2,j,k}^* + \frac{\delta t}{2 \delta y}
\left(F_{y,i,j-1/2,k}^{*|z} - F_{y,i,j+1/2,k}^{*|z} \right) \\
 & & + \frac{\delta t}{2 \delta z}
\left(F_{z,i,j,k-1/2}^{*|y} - F_{z,i,j,k+1/2}^{*|y} \right) \\
q_{Rx,i+1/2,j,k}^{n+1/2} &=& q_{Rx,i+1/2,j,k}^* + \frac{\delta t}{2 \delta y}
\left(F_{y,i+1,j-1/2,k}^{*|z} - F_{y,i+1,j+1/2,k}^{*|z} \right) \\
 & & + \frac{\delta t}{2\delta z}
\left(F_{z,i+1,j,k-1/2}^{*|y} - F_{z,i+1,j,k+1/2}^{*|y} \right) 
\end{eqnarray}
with $y$- and $z$-interface states being defined in an equivalent
manner by cyclic permutation of $(x,y,z)$ and $(i,j,k)$.  For each of
the $\delta t/2$ updated interface states, calculate the associated
flux, giving the $x$-interface flux
\beq
F_{x,i+1/2,j,k}^{n+1/2} = \nf_x(q_{Lx,i+1/2,j,k}^{n+1/2},q_{Rx,i+1/2,j,k}^{n+1/2})
\eeq
and similar expressions for the $y$- and $z$-interface fluxes.

\par
Step 4, update the the conserved variables from time $n$ to $n+1$ via
the fully corner coupled numerical fluxes
\begin{eqnarray}
q_{i,j,k}^{n+1} &=& q_{i,j,k}^{n}
 + \frac{\delta t}{\delta x}
\left(F_{x,i-1/2,j,k}^{n+1/2} - F_{x,i+1/2,j,k}^{n+1/2} \right) \\
 & & 
 + \frac{\delta t}{\delta y}
\left(F_{y,i,j-1/2,k}^{n+1/2} - F_{y,i,j+1/2,k}^{n+1/2} \right) 
 + \frac{\delta t}{\delta z}
\left(F_{z,i,j,k-1/2}^{n+1/2} - F_{z,i,j,k+1/2}^{n+1/2} \right) .
\end{eqnarray}
This completes the description of the 12-solve CTU algorithm for a
typical system of conservation laws, such as Euler's equations.
Unfortunately, as written above, the 12-solve CTU algorithm does not
result in a useful method for ideal MHD.  This can be understood on
rather general grounds by noting that the intermediate steps in the
algorithm use partial updates based on a dimensional splitting of the
equations in conservation form.  This in turn ignores the potential
balance between flux gradients in different directions (in particular
MHD source terms associated with those flux gradients) which is always
present for MHD owing to the ${\bf\nabla\cdot B}=0$ constraint.  We
make this point more concrete by considering two points in detail.

\par
First, note that the parallel flux gradient terms ($x$-flux gradient
at $x$-interfaces, etc.) are included in the PPM interface states
algorithm using the dimensionally split, primitive form of the
equations for MHD.  Meanwhile, the transverse flux gradient terms are
included using the conservative form of the equations.  Since the
dimensionally split primitive and conservative form of the equations
for MHD are not commensurate, this amounts to neglecting certain MHD
source terms resulting in a formally first order accurate integration
algorithm.  In addition, such an algorithm would also show secular
evolution of a magnetic field component perpendicular to the magnetic
field loop in the \emph{gedanken} experiment discussed in
{}\S\ref{sec:lr_states}.  Note that the same is true of the
two-dimensional CTU algorithm \cite{GS05}, which required the addition
of source terms to the transverse flux gradient correction step.  For
the 3D 12-solve CTU algorithm, with two predictor steps, this source
term correction procedure is increasingly complicated.

\par
To see why this is so, consider step 2 in the description of the
12-solve CTU algorithm just presented.  In particular, note that in
this step one generates two interface states at each interface,
by evolving the PPM interface state by $\delta t/3$ of one
transverse flux gradient.  In other words, for each interface normal
component of the magnetic field (used to define the divergence of the
magnetic field), one generates two normal components, each of
which is evolved by one half of a CT or Stokes EMF field loop.  (As an
aside, note that at this stage ${\bf\nabla\cdot B}=0$ is satisfied in
the sense of the average of these normal magnetic field components.)
It is useful at this stage to consider again the {}\emph{gedanken}
experiment discussed in {}\S\ref{sec:lr_states} and consider the
generation of the two $z$-interface states in step 2 of the
integration algorithm.  As a result of the MHD source terms in
the $x$- and $y$-flux gradients, namely $v_z \left(\partial B_x /
\partial x\right)$ and $v_z \left(\partial B_y / \partial y\right)$,
these two $z$-interface states will show a non-zero, in fact equal and
opposite, evolution of $B_z$.  This is a manifestation of a failure to
satisfy the balance condition discussed in {}\S\ref{sec:lr_states}.
Thus what we find is that it is the balanced, dimensionally split
system presented in {}\S\ref{sec:lr_states}, equations
(\ref{eq:Bx_induction_lim})-(\ref{eq:Bz_induction_lim}), which should
replace the straight forward dimensionally split conservative form of
the induction equation in steps 2 and 3 of the 12-solve CTU algorithm.
In practice, this means that the interface \emph{normal} components of
the magnetic field must incorporate a source term so as to maintain
this balance, and the two, say $z$-interface states, which are
generated must use the same source term, with opposite sign, so as to
maintain the magnetic divergence condition in an average sense.
Finally, note that as a result of dimensionally splitting the system,
the momentum and energy update in step 2 and 3 also require the
addition of source terms to balance terms like $B_x
\left(\partial B_x / \partial x\right)$ and ${\bf B\cdot v}
\left(\partial B_x / \partial x\right)$ etc.

\par
The net result is that a well balanced, 12-solve CTU algorithm for
ideal MHD can be constructed by using partial updates based on a
dimensional splitting of the MHD equations using a carefully chosen,
non-conservative form for the intermediate steps.  This form is found
by paying particular attention to the implicit balance between the
flux gradients, as was done for the induction equation in
{}\S\ref{sec:lr_states} for the PPM interface state algorithm.  The
advantage of this approach is a computational algorithm which is
optimally stable for CFL numbers $\leq 1$.  The disadvantage is that
the algorithm is complicated.  We have implemented the 12-solve MHD
CTU algorithm as described above and present results of tests of the
method in {}\S\ref{sec:tests}.  However, the complexity of the method
motivates us to find a simpler alternative, which we describe below.

\subsection{6-solve CTU variant for MHD}
\label{sec:6_solve}

In this subsection we present a simple variant on the 12-solve CTU
algorithm which we will henceforth refer to as the 6-solve algorithm.
For Euler's equations, the 6-solve algorithm can be described
concisely as the 12-solve CTU algorithm of \S\ref{sec:12_solve} omitting
step 2 and replacing $F_{x,i+1/2,j,k}^{*|y}$ and $F_{x,i+1/2,j,k}^{*|z}$
with $F_{x,i+1/2,j,k}^*$ (and similarly for the $y$- and $z$-fluxes) in
step 3.  Alternatively, one may also describe it as a formal extension of
the 2D CTU algorithm in which the parallel and transverse flux gradients
are included in the interface states in a two-step process.  In what
follows we present a functional description of this 6-solve CTU algorithm
for MHD including a detailed description of the treatment of the MHD
source terms and constrained transport electric fields.

\par
Step 1, calculate the left and right PPM interface states
$q_{Lx,i+1/2,j,k}^*$, $q_{Rx,i+1/2,j,k}^*$, $q_{Ly,i,j+1/2,k}^*$,
$q_{Ry,i,j+1/2,k}^*$, $q_{Lz,i,j,k+1/2}^*$ and $q_{Rz,i,j,k+1/2}^*$
including the MHD source terms described in \S\ref{sec:3D_lr_states}
and the associated interface fluxes
\begin{eqnarray}
F_{x,i+1/2,j,k}^* &=& \nf_x(q_{Lx,i+1/2,j,k}^*,q_{Rx,i+1/2,j,k}^*) \\ 
F_{y,i,j+1/2,k}^* &=& \nf_y(q_{Ly,i,j+1/2,k}^*,q_{Ry,i,j+1/2,k}^*) \\ 
F_{z,i,j,k+1/2}^* &=& \nf_z(q_{Lz,i,j,k+1/2}^*,q_{Rz,i,j,k+1/2}^*) .
\end{eqnarray}

\par
Step 2, apply the CT algorithm of \S\ref{sec:CT_Alg} to calculate the CT
electric fields $\E_{x,i,j+1/2,k+1/2}^*$, $\E_{y,i+1/2,j,k+1/2}^*$
and $\E_{z,i+1/2,j+1/2,k}^*$ using the numerical fluxes from step 1
and a cell center reference electric field calculated using the
initial data at time level $n$, i.e. $q_{i,j,k}^{n}$.

\par
Step 3, at each interface evolve the PPM interface states by $\delta
t/2$ of the transverse flux gradients.  The hydrodynamic variables
(mass, momentum and energy density) are advanced using
\begin{eqnarray}
q_{Lx,i+1/2,j,k}^{n+1/2} &=& q_{Lx,i+1/2,j,k}^* + \frac{\delta t}{2 \delta y}
\left(F_{y,i,j-1/2,k}^* - F_{y,i,j+1/2,k}^* \right) \nonumber \\
 & & + \frac{\delta t}{2 \delta z}
\left(F_{z,i,j,k-1/2}^* - F_{z,i,j,k+1/2}^* \right) +
\frac{\delta t}{2} S_{x,i,j,k} \\
q_{Rx,i+1/2,j,k}^{n+1/2} &=& q_{Rx,i+1/2,j,k}^* + \frac{\delta t}{2 \delta y}
\left(F_{y,i+1,j-1/2,k}^* - F_{y,i+1,j+1/2,k}^* \right) \nonumber \\
 & & + \frac{\delta t}{2\delta z}
\left(F_{z,i+1,j,k-1/2}^* - F_{z,i+1,j,k+1/2}^* \right) +
\frac{\delta t}{2} S_{x,i+1,j,k}
\end{eqnarray}
where the $x$-interface MHD source term for the momentum density
\beq
(S_{x,i,j,k})_{\rho{\bf v}} = {\bf B}_{i,j,k}
\left( \frac{\partial B_x}{\partial x} \right)_{i,j,k}
\eeq
and the energy density
\begin{eqnarray}
(S_{x,i,j,k})_{E} & = & (B_y v_y)_{i,j,k} 
\mm{-\frac{\partial B_z}{\partial z}}{\frac{\partial B_x}{\partial x}}_{i,j,k} +
\nonumber \\
 & & (B_z v_z)_{i,j,k} 
\mm{-\frac{\partial B_y}{\partial y}}{\frac{\partial B_x}{\partial x}}_{i,j,k}
~.
\end{eqnarray}
The magnetic field components are evolved using the CT electric fields in
place of the predictor fluxes.  The interface normal component of the
magnetic field is evolved using the integral form of the Stokes loop,
\begin{eqnarray}
B_{x,i+1/2,j,k}^{n+1/2} & = & B_{x,i+1/2,j,k}^n - \frac{\delta t}{2 \delta y}
\left(\E_{z,i+1/2,j+1/2,k}^* - \E_{z,i+1/2,j-1/2,k}^* \right) \nonumber \\
 & & + \frac{\delta t}{2 \delta z}
\left(\E_{y,i+1/2,j,k+1/2}^* - \E_{y,i+1/2,j,k-1/2}^* \right) ~.
\end{eqnarray}
The $y$-component of the magnetic field is evolved using
\begin{eqnarray}
(B_y)_{Lx,i+1/2,j,k}^{n+1/2} &=& (B_y)_{Lx,i+1/2,j,k}^* -
 \frac{\delta t}{4 \delta z}
\left(\E_{x,i,j+1/2,k+1/2}^* - \E_{x,i,j+1/2,k-1/2}^* \right) \nonumber \\
 & & -\frac{\delta t}{4 \delta z}
\left(\E_{x,i,j-1/2,k+1/2}^* - \E_{x,i,j-1/2,k-1/2}^* \right) \nonumber \\
 & & + \frac{\delta t}{2} (S_{x,i,j,k})_{B_y} \\
(B_y)_{Rx,i+1/2,j,k}^{n+1/2} &=& (B_y)_{Rx,i+1/2,j,k}^* -
 \frac{\delta t}{4 \delta z}
\left(\E_{x,i+1,j+1/2,k+1/2}^* - \E_{x,i+1,j+1/2,k-1/2}^* \right) \nonumber \\
 & & -\frac{\delta t}{4 \delta z} 
\left(\E_{x,i+1,j-1/2,k+1/2}^* - \E_{x,i+1,j-1/2,k-1/2}^* \right) \nonumber \\
 & & + \frac{\delta t}{2} (S_{x,i+1,j,k})_{B_y}
\end{eqnarray}
with
\beq
(S_{x,i,j,k})_{B_y} = (v_y)_{i,j,k}
\mm{-\frac{\partial B_z}{\partial z}}{\frac{\partial B_x}{\partial x}}_{i,j,k}
~.
\eeq
The $z$-component of the magnetic field is evolved using
\begin{eqnarray}
(B_z)_{Lx,i+1/2,j,k}^{n+1/2} &=& (B_z)_{Lx,i+1/2,j,k}^* +
 \frac{\delta t}{4 \delta y}
\left(\E_{x,i,j+1/2,k+1/2}^* - \E_{x,i,j-1/2,k+1/2}^* \right) \nonumber \\
 & & + \frac{\delta t}{4 \delta y}
\left(\E_{x,i,j+1/2,k-1/2}^* - \E_{x,i,j-1/2,k-1/2}^* \right) \nonumber \\
 & & + \frac{\delta t}{2} (S_{x,i,j,k})_{B_z} \\
(B_z)_{Rx,i+1/2,j,k}^{n+1/2} &=& (B_z)_{Rx,i+1/2,j,k}^* +
 \frac{\delta t}{4 \delta y}
\left(\E_{x,i+1,j+1/2,k+1/2}^* - \E_{x,i+1,j-1/2,k+1/2}^* \right) \nonumber \\
 & & + \frac{\delta t}{4 \delta y}
\left(\E_{x,i+1,j+1/2,k-1/2}^* - \E_{x,i+1,j-1/2,k-1/2}^* \right) \nonumber \\
 & & + \frac{\delta t}{2} (S_{x,i+1,j,k})_{B_z}
\end{eqnarray}
with
\beq
(S_{x,i,j,k})_{B_z} = (v_z)_{i,j,k}
\mm{-\frac{\partial B_y}{\partial y}}{\frac{\partial B_x}{\partial x}}_{i,j,k} 
~.
\eeq
Note that the origin of these MHD source terms for the transverse
components of the magnetic field can be clearly seen as resulting from
the directional splitting of the induction equation described in
\S\ref{sec:3D_lr_states}.  The momentum and energy density MHD source
terms originate from the use of the primitive variable form of the MHD
equations to calculate the PPM interface states.  The $y$- and
$z$-interface states are advanced in an equivalent manner by cyclic
permutation of $(x,y,z)$ and $(i,j,k)$ in the above expressions.

\par
Step 4, for each of the $\delta t/2$ updated interface states, calculate
the associated flux, giving the $x$-interface flux
\beq
F_{x,i+1/2,j,k}^{n+1/2} = \nf_x(q_{Lx,i+1/2,j,k}^{n+1/2},q_{Rx,i+1/2,j,k}^{n+1/2}) 
\eeq
and similar expressions for the $y$- and $z$-interface fluxes.

\par
Step 5, apply the CT algorithm of \S\ref{sec:CT_Alg} to calculate the
CT electric fields $\E_{x,i,j+1/2,k+1/2}^{n+1/2}$,
$\E_{y,i+1/2,j,k+1/2}^{n+1/2}$ and $\E_{z,i+1/2,j+1/2,k}^{n+1/2}$
using the numerical fluxes from step 4 and a cell center reference
electric field calculated using the cell average state at time level
$n+1/2$ which is calculated as follows.  The cell center magnetic
field components are defined as equaling the arithmetic average of the
interface magnetic field components, $B_{x,i,j,k}^{n+1/2} =
(B_{x,i+1/2,j,k}^{n+1/2} + B_{x,i-1/2,j,k}^{n+1/2})/2$ and similarly
for the $y$- and $z$-components.  The mass and momentum density are
computed using a conservative update with the predictor fluxes from
step 1,
\begin{eqnarray}
q_{i,j,k}^{n+1/2} &=& q_{i,j,k}^{n}
 + \frac{\delta t}{2 \delta x}
\left(F_{x,i-1/2,j,k}^* - F_{x,i+1/2,j,k}^* \right) \\
& + & \frac{\delta t}{2 \delta y}
\left(F_{y,i,j-1/2,k}^* - F_{y,i,j+1/2,k}^* \right) 
+ \frac{\delta t}{2 \delta z}
\left(F_{z,i,j,k-1/2}^* - F_{z,i,j,k+1/2}^* \right) ~.
\nonumber
\end{eqnarray}

\par
Step 6, update the solution from time level $n$ to $n+1$.  The
hydrodynamic variables (mass, momentum and energy density) are
advanced using the standard the flux integral relation,
\begin{eqnarray} 
q_{i,j,k}^{n+1} &=& q_{i,j,k}^{n}
 + \frac{\delta t}{\delta x}
\left(F_{x,i-1/2,j,k}^{n+1/2} - F_{x,i+1/2,j,k}^{n+1/2} \right) \\
 & & + \frac{\delta t}{\delta y}
\left(F_{y,i,j-1/2,k}^{n+1/2} - F_{y,i,j+1/2,k}^{n+1/2} \right)
+ \frac{\delta t}{\delta z}
\left(F_{z,i,j,k-1/2}^{n+1/2} - F_{z,i,j,k+1/2}^{n+1/2} \right) 
\nonumber
\end{eqnarray}
and the interface averaged normal components of the magnetic field are
advanced using a Stokes loop integral, 
\begin{eqnarray}
B_{x,i+1/2,j,k}^{n+1} & = & B_{x,i+1/2,j,k}^n - \frac{\delta t}{\delta y}
\left(\E_{z,i+1/2,j+1/2,k}^{n+1/2} - \E_{z,i+1/2,j-1/2,k}^{n+1/2} \right)
\nonumber \\
 & & + \frac{\delta t}{\delta z}
\left(\E_{y,i+1/2,j,k+1/2}^{n+1/2} - \E_{y,i+1/2,j,k-1/2}^{n+1/2} \right) ~,
\end{eqnarray}
\begin{eqnarray}
B_{y,i,j+1/2,k}^{n+1} & = & B_{y,i,j+1/2,k}^n + \frac{\delta t}{\delta x}
\left(\E_{z,i+1/2,j+1/2,k}^{n+1/2} - \E_{z,i-1/2,j+1/2,k}^{n+1/2} \right)
\nonumber \\
 & & - \frac{\delta t}{\delta z}
\left(\E_{x,i,j+1/2,k+1/2}^{n+1/2} - \E_{x,i,j+1/2,k-1/2}^{n+1/2} \right) ~,
\end{eqnarray}
and
\begin{eqnarray}
B_{z,i,j,k+1/2}^{n+1} & = & B_{z,i,j,k+1/2}^n - \frac{\delta t}{\delta x}
\left(\E_{y,i+1/2,j,k+1/2}^{n+1/2} - \E_{y,i-1/2,j,k+1/2}^{n+1/2} \right)
\nonumber \\
 & & + \frac{\delta t}{\delta y}
\left(\E_{x,i,j+1/2,k+1/2}^{n+1/2} - \E_{x,i,j-1/2,k+1/2}^{n+1/2} \right) ~.
\end{eqnarray}

\par
This completes the description of the 6-solve CTU algorithm.  This
relatively simple 3D integration algorithm is second order accurate
and has the advantage over the 12-solve CTU algorithm that no source
terms need be included in the evolution of the interface normal
components of the magnetic field.  This algorithm is designed in such
a way that for grid aligned flows it reduces exactly to the 2D CTU and
1D PPM integration algorithms for problems involving the relevant
symmetry.  Additionally, consideration of the field loop advection
\emph{gedanken} experiment described in {}\S\ref{sec:lr_states} shows
that the 6-solve CTU algorithm is well balanced and preserves the
$B_z=0$ condition exactly.  The downside of the present 6-solve
algorithm is that we observe experimentally that the algorithm is
stable for CFL $< 1/2$.  When compared to the 12-solve algorithm, this
is compensated by the fact that it requires half as many Riemann
solutions per time step.  Hence, to a large extent the 6-solve and
12-solve algorithms show similar computational cost: two time-steps
with the 6-solve algorithm at a CFL number of 1/2 is nearly equivalent
to one time-step with the 12-solve algorithm with a CFL number of one.

\section{Tests}
\label{sec:tests}

In this section we present results obtained with the 6-solve CTU + CT
integration algorithm just described.  For the sake of comparison, and
clarification of the dominant differences between the results obtained
with the 6-solve and 12-solve algorithms, some results using the
12-solve algorithm will also be included.  We will find in this
section, through a series of tests, that the dominant difference
between the two are the stability domain.  Otherwise, they are quite
comparable in accuracy and computational cost.  As a result, in
practical applications we prefer the 6-solve algorithm for MHD on
account of its simplicity and smaller memory footprint.

\subsection{Field Loop Advection}
\label{subsec:fieldloop}

In this section we discuss and present results for the advection of a
magnetic field loop.  In order to narrow the focus and clarify the
discussion, in this section we will concern ourselves primarily with
the the initial conditions in which the density $\rho$, velocity ${\bf
v}$, and pressure $P$ are constants, and the magnetic field is weak in
the sense that $\beta = 2P/B^2 \gg 1$.  In this limit, the evolution
equations for the magnetic field are well approximated by the
advection of a set of passive scalar functions, say the components of
the magnetic vector potential.  In the construction of the 2D CTU-CT
algorithm {}\cite{GS05} as well as the 3D 6-solve and 12-solve
algorithms presented here we have found that recovering the correct
solution in this limiting case can be surprisingly difficult for
conservative, finite volume algorithms applied to the ideal MHD
equations.

\par
As a concrete example of a situation in which this problem can be
challenging, consider the field loop advection test problem studied in
{}\cite{GS05} and discussed as a \emph{gedanken} experiment in
{}\S\ref{sec:lr_states}.  Specifically, consider a field loop confined
to the $(x,y)$-plane, i.e. $B_z=0$, and a constant advection velocity
field with $v_z \neq 0$.  If care is not taken to respect the balance
between the MHD source terms in calculating the interface states,
updating them with transverse flux gradients, etc. one can find an
erroneous and sometimes \emph{secular} evolution of $B_z$.  It should
be noted that these concerns are not limited to CTU or PPM which use a
predictor step.  For example schemes with are conservative but do not
satisfy ${\bf\nabla\cdot B}=0$ will also find erroneous evolution for
$B_z$ in this problem.  There is an interesting corollary which
results from this observation.  If a conservative numerical algorithm
can solve this magnetic field loop advection problem and preserve the
solution $B_z=0$ for all time, it also satisfies the ${\bf\nabla\cdot
B}=0$ condition.

\par
We begin by selecting a computational domain $-0.5\le x \le 0.5$, $-0.5
\le y \le 0.5$, and $-1 \le z \le 1$, resolved on a $N \times
N \times 2N$ grid and apply periodic boundary conditions.  The
hydrodynamical state is uniform with a density $\rho=1$, pressure
$P=1$, and velocity components $(v_x,~v_y,~v_z) = (1,~1,~2)$.  The
initialization of the magnetic field is most easily described in terms
of a vector potential in the coordinate system $(x_1,~x_2,~x_3)$ which
is related to the computational coordinate system $(x,~y,~z)$ via the
rotation
\begin{eqnarray}
x_1 & = & (2 x + z)/\sqrt{5}
\nonumber \\
x_2 & = & y
\nonumber \\
x_3 & = & (-x + 2 z)/\sqrt{5} ~.
\label{eq:loop_coord_trans}
\end{eqnarray}
In particular, we choose $A_1 = A_2 = 0$ and 
\beq
A_3 = \left \{
\begin{array}{ll}
B_0 (R - r) & \textrm{for}~ r \le R \\
0           & \textrm{for}~ r > R
\end{array}
\right .
\eeq
where $B_0=10^{-3}$, $R=0.3$ and $r=\sqrt{x_1^2+x_2^2}$ in the domain
$-0.5\lambda_1 \le x_1 \le 0.5\lambda_1$, $-0.5 \lambda_2 \le x_2 \le
0.5 \lambda_2$.  To satisfy the periodic boundary conditions we choose
$\lambda_1 = 2/\sqrt{5}$ and $\lambda_2 = 1$ and define $A_3(x_1 +
n\lambda_1, x_2 + m \lambda_2, x_3) = A_3(x_1, x_2, x_3)$ for all
integers $(n,~m)$.

\par
\begin{figure}[hbt]
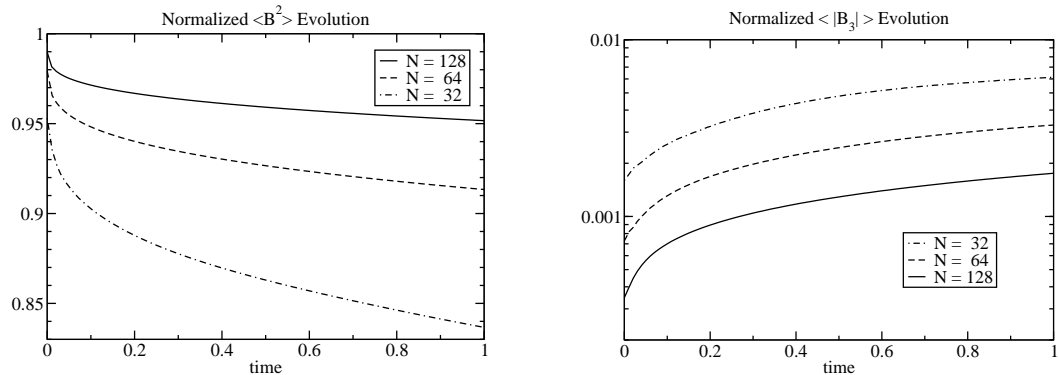

\begin{center}
\includegraphics*[width=2.5in]{LoopMagEnergy.eps} \hfill
\includegraphics*[width=2.5in]{LoopB3Energy.eps}
\end{center}
\caption{Time evolution of the normalized, volume average magnetic
energy density $<B^2>$ and the component along the $x_3$-direction,
$<B_3^2>$, for three different grid resolutions using the 6-solve
integration algorithm.}
\label{fig:loop_energy}
\end{figure}

As a quantitative measure of the dissipation in the algorithm we plot
the time evolution of the volume averaged magnetic energy density
$<B^2>$ normalized to the initial (analytic) value $<B^2> =
(\sqrt{5}\pi R^2/2) B_0^2$ in Figure \ref{fig:loop_energy}.
Interestingly, the magnetic energy density \mbox{$<B^2>$} shows a
temporal evolution similar to what was observed with the 2D algorithm
in \cite{GS05}.  Namely, it can be well fit as a power law of the form
\mbox{$<B^2>=C(1-(t/\tau)^\alpha)$} where $\tau =
(3.22\times10^2,~3.68\times10^3,~2.65\times10^4)$ and $\alpha =
(0.365,~0.328,~0.320)$ for $N=(32,~64,~128)$ respectively.  Moreover,
these values are quite comparable to the time constant
$\tau=1.061\times10^4$ and exponent $\alpha=0.291$ found in the 2D
calculation.

\par
For the specific case of a cylindrical magnetic field loop with
translation invariance in the $z$-direction $(\partial/\partial z =
0)$ the 6-solve and 12-solve algorithms studied in this paper reduce
exactly to the 2D algorithm presented in {}\cite{GS05}.  As such, the
3D algorithms presented here give the same solution as the 2D
algorithm in {}\cite{GS05} and preserve the solution $B_z=0$ for all
time (we have explicitly tested that this is true with our
implementation of the method).  With the axis of the cylindrical field
loop is aligned along a non-special direction with respect to the
grid, preserving this property is non-trivial.

\par
\begin{figure}[hbt]
\begin{center}
\includegraphics*[width=2.5in]{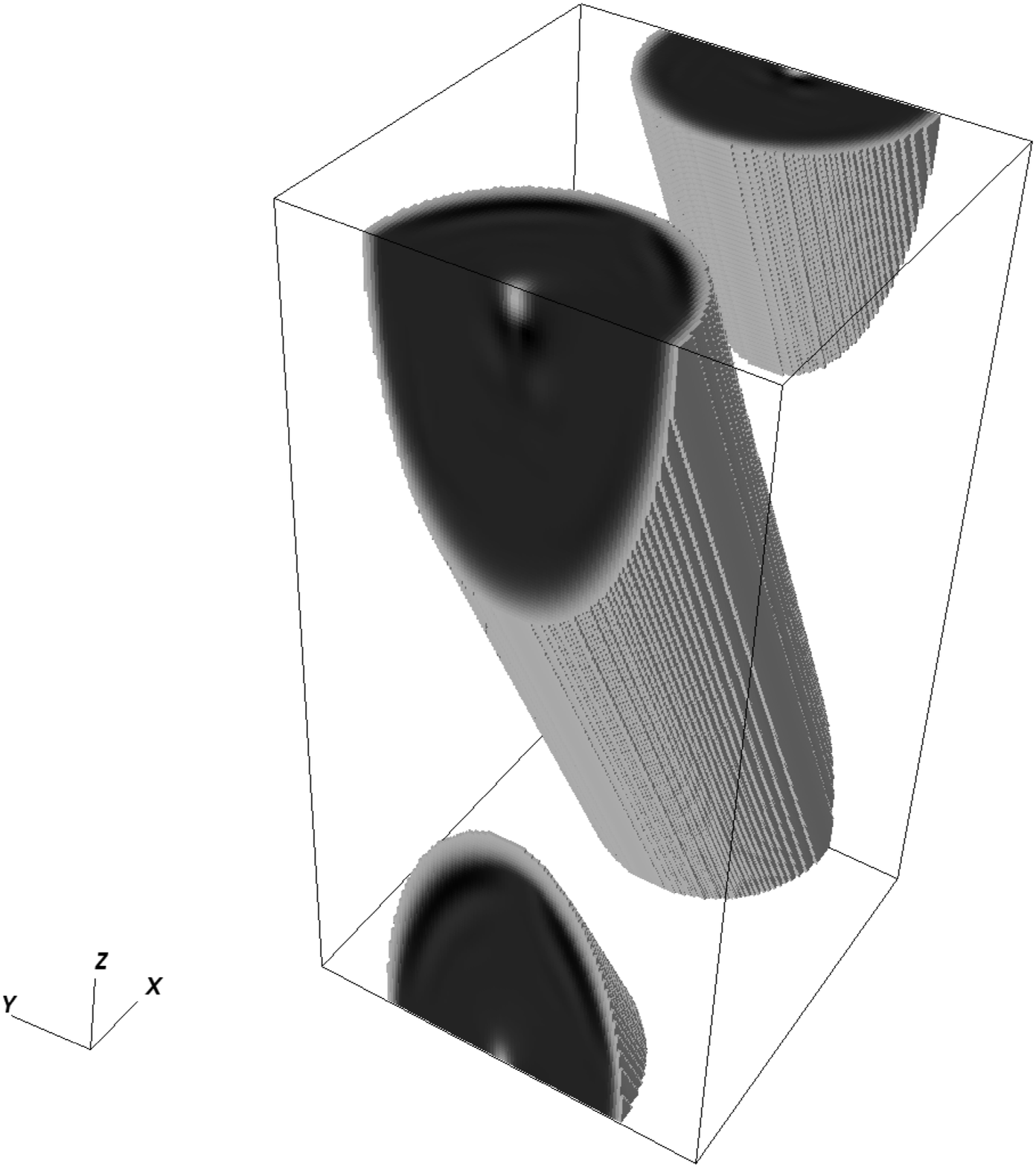} \hfill
\includegraphics*[width=2.5in]{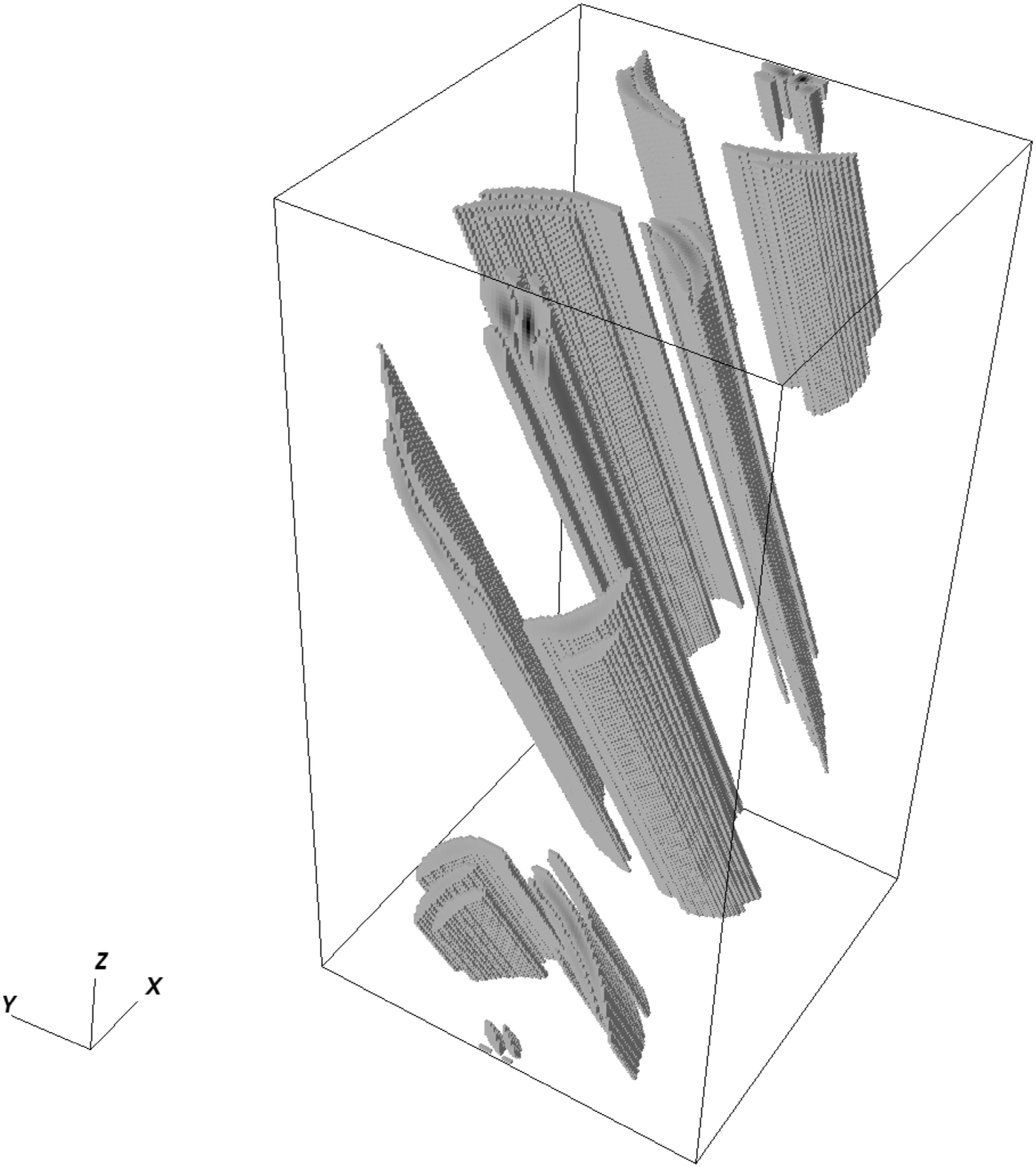}
\end{center}
\caption{Thresholded image of the magnetic energy ({\it left}) and the
$3$-component of the magnetic energy, $B_3^2/2$, at time = 1.}
\label{fig:loop_3d_images}
\end{figure}

As a quantitative measure of the ability of the algorithm to preserve
$B_3=0$ we plot the normalized error $<|B_3|>/B_0$ in Figure
{}\ref{fig:loop_energy}.  This error is calculated by contracting the
cell center magnetic field with a unit vector in the $x_3$-direction
and computing the volume average of its absolute value.  From this
plot it is clear that the convergence rate of $<|B_3|>/B_0$ as
measured in either the initial conditions, or the solution at time
$=1$ is approximately first order.  This behavior is consistent with
the observation that the $3$-component of the magnetic energy is
dominant on the axis and at the boundary of the magnetic cylinder
where the current density is initially singular, as shown in figure
{}\ref{fig:loop_3d_images}.  It is also worth noting that away from
these regions, the solution preserves the 3-component of the magnetic
energy quite small.  This would not be the case if care were not taken
to balance the MHD source terms in the integration algorithm.

%

\subsection{Linear Wave Convergence}
\label{sec:linear_wave}

In this subsection we present the results of a convergence study for
both the 6-solve and 12-solve CTU-CT MHD algorithms.  The problem we
study is the propagation of linear amplitude, planar waves in a
direction which is oblique to the grid.  The physical conditions of
the problem are most easily described in a coordinate system
$(x_1,~x_2,~x_3)$ which is chosen such that the wave propagates
parallel to the $x_1$-axis.  In this coordinate system, the initial
conserved variable state vector is given by
\beq
q^0 = \bar{q} + \varepsilon R_p \cos\left(\frac{2 \pi x_1}{\lambda}\right)
\eeq
where $\bar{q}$ is the mean background state, $\varepsilon=10^{-6}$ is
the wave amplitude, and $R_p$ is the right eigenvector in conserved
variables for wave mode $p$ (calculated in the state $\bar{q}$).  In
order to enable others to perform the same tests presented here and
compare the results in a quantitative manner, we include the numerical
values for the right eigenvectors in the appendix.

\par
The mean background state $\bar{q}$ is selected so that the wave
speeds are well separated and there are no inherent symmetries in the
magnetic field orientation (when initialized on the grid).  The
density $\bar{\rho}=1$ and gas pressure $\bar{P}=1/\gamma=3/5$.  The
velocity component $\bar{v}_1=1$ for the entropy mode test and
$\bar{v}_1=0$ for all other wave modes.  The transverse velocity
components $\bar{v}_2=\bar{v}_3=0$.  The magnetic field components
$\bar{B}_1=1$, $\bar{B}_2=3/2$, and $\bar{B}_3=0$.  With this choice,
the slow mode speed $c_s=1/2$, the Alfv\'en speed $c_a=1$, and the
fast mode speed $c_f=2$ in the $x_1$-direction.

\par
The computational domain extends from $0\le x \le 3.0$, $0 \le y \le
1.5$, and $0 \le z \le 1.5$, is resolved on a $2N \times N \times N$
grid and uses periodic boundary conditions.  Initializing this problem
on the computational grid is accomplished by applying a coordinate
transformation
\begin{eqnarray}
x & = & x_1 \cos\alpha \cos\beta - x_2 \sin\beta - x_3 \sin\alpha \cos\beta 
\nonumber \\
y & = & x_1 \cos\alpha \sin\beta + x_2 \cos\beta - x_3 \sin\alpha \sin\beta 
\nonumber \\
z & = & x_1 \sin\alpha                           + x_3 \cos\alpha
\label{eq:wave_coord_trans}
\end{eqnarray}
from the $(x_1,~x_2,~x_3)$ coordinate system to the $(x,~y,~z)$
coordinate system of the grid with $\sin\alpha = 2/3$ and $\sin\beta =
2/\sqrt{5}$.  With this choice, there is one wave period along each
grid direction and the wavelength $\lambda=1$.  The interface
components of the magnetic field are initialized via a magnetic vector
potential so as to ensure ${\bf\nabla\cdot B}=0$.

\par
\begin{figure}[hbt]
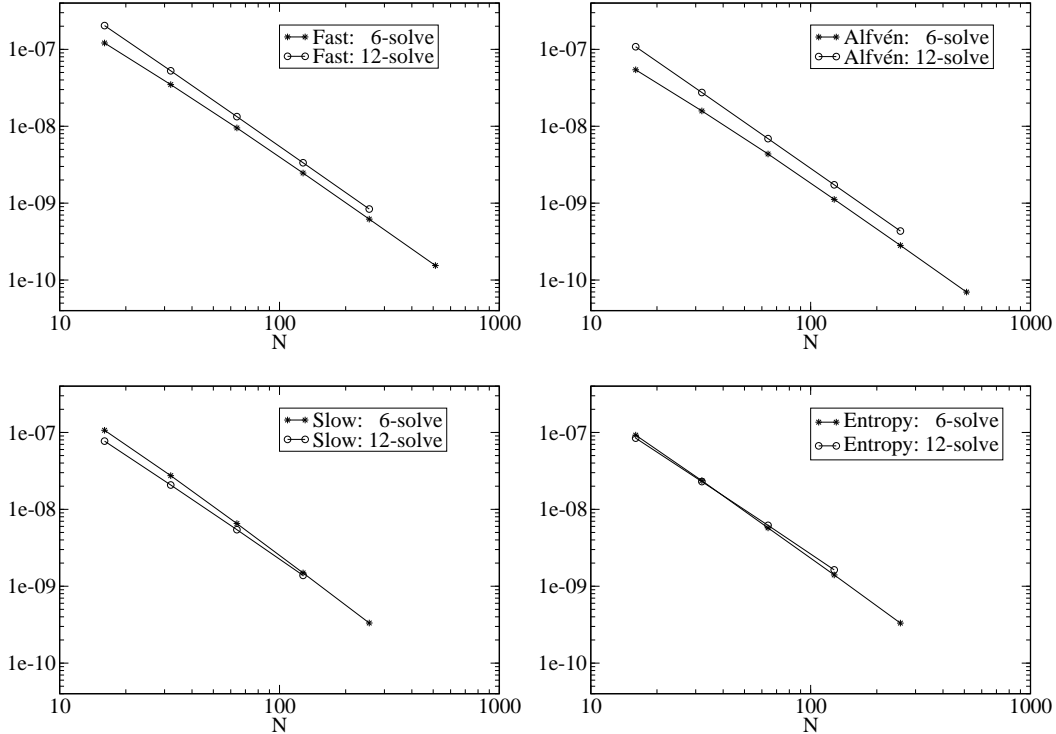

\begin{center}
\includegraphics*[width=2.7in]{L1error.fast_bw.eps} \hfill
\includegraphics*[width=2.7in]{L1error.alfven_bw.eps} \\[12pt]
\includegraphics*[width=2.7in]{L1error.slow_bw.eps} \hfill
\includegraphics*[width=2.7in]{L1error.entropy_bw.eps}
\end{center}
\caption{Linear wave convergence of fast, Alfv\'en, slow and entropy modes
using the CTU + CT 6-solve and 12-solve integration algorithms.  The
symbols denote the calculated L1-error norm.}
\label{fig:LinWaveError}
\end{figure}

The error in the solution is calculated after propagating the wave for
a distance equal to one wavelength.  Hence, the initial state is
evolved for a time $t=\lambda/c$ where $c$ is the speed of the wave
mode under consideration.  For each component $s$ of the conserved
variable vector $q$ we calculate the L1-error with respect to the
initial conditions
\beq
\delta q_s = \frac{1}{2N^3} \sum_{i,j,k} |q_{i,j,k,s}^n - q_{i,j,k,s}^0|
\eeq
by summing over all grid cells $(i,j,k)$.  We use the cell center
components of the magnetic field in computing this error.  In figure
\ref{fig:LinWaveError} we plot the norm of this error vector
\beq
\| \delta q \| = \sqrt{ \sum_s (\delta q_s)^2 } 
\label{eq:error_norm}
\eeq
for the fast, Alfv\'en, slow and entropy modes.  Both algorithms
demonstrate a second order convergence.  With the exception of the
slow mode, the 6-solve algorithm shows lower errors than the 12-solve
algorithm.  Note that the choice of maximum resolution in the
convergence study for each algorithm and wave mode was selected on the
basis of the ``cost'' of the computation.

%

\subsection{Circularly Polarized Alfv\'en Wave}
\label{sec:cp_alfven}

In this section we present results for the propagation of a circularly
polarized Alfv\'en wave in a periodic domain using both the 6-solve
and 12-solve algorithms.  This problem is interesting from the
perspective that the wave is an exact nonlinear solution to the ideal
MHD equations.  Hence, this problem enables one to easily measure the
nonlinear convergence to a multidimensional solution of the ideal MHD
equations \cite{Toth-divB}.  This problem is also interesting from the
point of view that for a range of parameters, the circularly polarized
Alfv\'en wave is susceptible to a parametric instability
{}\cite{Goldstein,DelZanna-ea-01}.  Unfortunately, this situation has
also hindered its applicability as a general and robust test for
multidimensional MHD algorithms
{}\cite{Pen-MHD-03,Londrillo-DelZanna04}.  For the parameters used
here, and suggested by T\'oth \cite{Toth-divB} we find no indication
of instability.

\par
As with the linear wave propagation study presented in
{}\S\ref{sec:linear_wave}, the initial conditions are most easily
described in a coordinate system $(x_1,~x_2,~x_3)$ which is chosen
such that the wave propagates parallel to the $x_1$-axis.  In this
coordinate system, the magnetic field components $B_1 = 1$, $B_2 =
0.1\sin(2\pi x_1/\lambda)$, and $B_3 = 0.1\cos(2\pi x_1/\lambda)$.
The velocity components $v_1 = (0,1)$ for traveling or standing
Alfv\'en waves respectively, $v_2 = 0.1\sin(2\pi x_1/\lambda)$, and
$v_3 = 0.1\cos(2\pi x_1/\lambda)$.  The mass density $\rho=1$ and the
gas pressure $P=0.1$, hence $\beta = 2P/B^2 \sim 0.2$.

\par
The computational domain used in this section is identical to that
used in {}\S\ref{sec:linear_wave}.  In particular, we use the
coordinate transformation given by equations \ref{eq:wave_coord_trans}
and a magnetic vector potential to initialize the magnetic fields so
as to ensure ${\bf\nabla\cdot B}=0$.  It is worth noting that this
approach will necessarily result in magnetic pressure perturbations as
a result of truncation error in initializing the magnetic field on the
grid.  The parallel component $B_1$ is a constant, and hence rotation
of this component will still result in a constant set of field
components with no ``pressure'' variation.  The perpendicular
components $(B_2,~B_3)$ however, will suffer some truncation error on
initialization.  Since $B_\perp^2/P = 0.1$ this truncation error in
initialization will drive compressive waves.  Note that with this set
of initial conditions and $v_1=0$ the Alfv\'en wave will travel a
distance of one wavelength $\lambda$ in a time $t=1$.

\par
\begin{figure}[hbt]
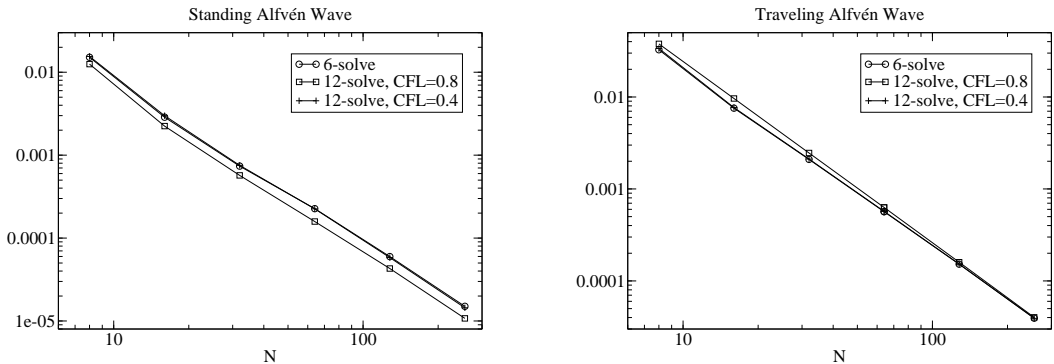

\begin{center}
\includegraphics*[width=2.5in]{cpaw_converge_stand.eps} \hfill
\includegraphics*[width=2.5in]{cpaw_converge_travel.eps}
\end{center}
\caption{L1 error norm for the 6-solve and 12-solve integration
algorithms for both standing (left) and traveling (right) circularly
polarized Alfv\'en waves. In particular note the dominant difference
between the 12-solve and 6-solve errors is attributable to the CFL
dependence.}
\label{fig:cpaw_converge}
\end{figure}

As a quantitative measure of the solution accuracy, we present in
figure \ref{fig:cpaw_converge} the norm of the L1 error vector (as
defined in equation {}\ref{eq:error_norm}) after propagating for a
time $t=1$ for both standing and traveling wave modes.  From this
figure we see that both traveling and standing circularly polarized
Alfv\'en waves converge with second order accuracy for both
integration algorithms.  The traveling wave mode shows a larger error
amplitude relative to the standing mode, but it is worth noting that
(while not shown here) the increase is fairly uniform over the
components of the error vector.  The 6-solve and 12-solve algorithms
show quite comparable errors for both standing and traveling wave
modes.  When using a CFL number of 0.4, the 12-solve and 6-solve
algorithms result in nearly identical errors.  Increasing the CFL
number to 0.8 with the 12-solve algorithm results in a slightly
reduced traveling wave error, and increased standing wave error.
These results indicate that the dominant difference in the L1 error
between the 6-solve and 12-solve algorithms results from the CFL
dependence of the truncation error.

\par
\begin{figure}[hbt]
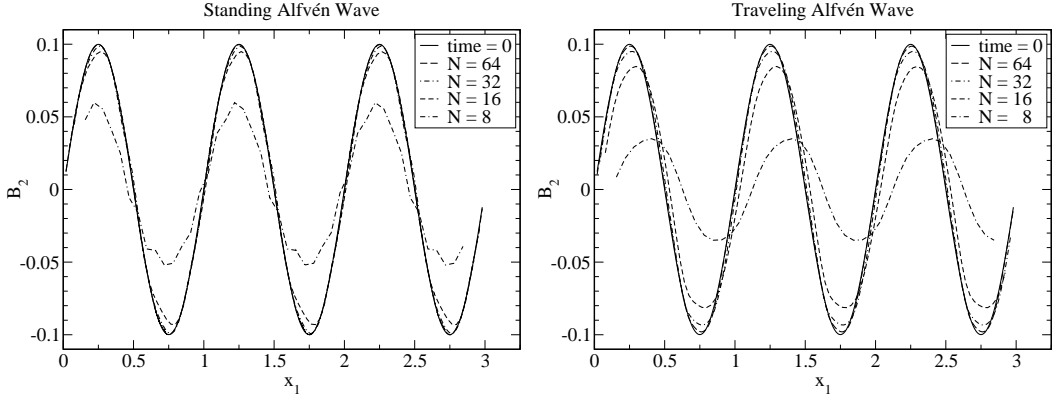

\begin{center}
\includegraphics*[width=2.7in]{cpaw3d_time5_v1.0.eps} \hfill
\includegraphics*[width=2.7in]{cpaw3d_time5_v0.0.eps}
\end{center}
\caption{Plot of $B_2$ versus $x_1$ at $t=5$ for the standing (left) and 
traveling (right) circularly polarized Alfv\'en waves using the
6-solve integration algorithm.  For comparison, the initial conditions
at $t=0$ for the $N=64$ case is also included.}
\label{fig:cpaw_time5}
\end{figure}

As a qualitative measure of the solution accuracy, we present in
figure \ref{fig:cpaw_time5} scatter plots of $B_2$ versus $x_1$ for
both standing and traveling wave modes after propagating for a time
$t=5$ using the 6-solve integration algorithm.  These plots are
constructed using the cell center components of the magnetic field,
the cell center position and the coordinate transformation given by
equations \ref{eq:wave_coord_trans}.  As a result of the fact that the
wave is rotated with respect to the grid, there are many grid cells
with the same cell center $x_1$-position.  Hence, since these plots
include every grid point in the grid, the lack of scatter in the plots
demonstrates that the Alfv\'en waves retain their planar symmetry
throughout the calculation.  Unfortunately, it is difficult to use the
results presented here to make direct contact with solutions presented
in the literature due to the scarcity of published three-dimensional
test solutions.  For analogous plots in a two-dimensional system see
{}\cite{Toth-divB,Pen-MHD-03,GS05}

\par
As a final measure of the solution accuracy and convergence, we
present results for the dissipation of magnetic helicity in the case
of a traveling circularly polarized Alfv\'en wave.  We note that this
is not the cleanest possible \emph{test}, since with periodic boundary
conditions and a mean magnetic field, it does not appear to be
generally possible to define a magnetic helicity which is conserved
{}\cite{Berger97}.  Nevertheless, we find that following
{}\cite{Brandenburg_Matthaeus_04} the magnetic helicity evolution
associated with the fluctuating components of the magnetic field gives
an interesting constraint on the problem considered here.  In
particular, let ${\bf B_0} = <{\bf B}>$ (where angle brackets denote a
spatial mean) and ${\bf b} = {\bf B} - {\bf B_0}$ denote the mean and
fluctuating components of the magnetic field respectively.  Also,
define the magnetic vector potential associated with the fluctuating
field as ${\bf b}=\nabla\times{\bf a}$.  It is worth noting that as a
result of periodic boundary conditions, ${\bf B_0}$ is time
independent and the magnetic helicity associated with the fluctuating
field $H = <{\bf b \cdot a}>$ is gauge invariant.  It follows that the
time evolution of the magnetic helicity is given by
\beq
\frac{d}{dt} <{\bf b}\cdot{\bf a}> = -2<{\bf E}\cdot{\bf b}>
\eeq
where ${\bf E}$ is the electric field.  Assuming ideal MHD, this
equation can also be written as
\beq
\frac{d}{dt} <{\bf b}\cdot{\bf a}> =
 -2 {\bf B_0} \cdot <{\bf v}\times{\bf b}> ~.
\eeq
From this expression it is clear for a circularly polarized Alfv\'en
wave the magnetic helicity should be conserved with $<{\bf b} \cdot
{\bf a}> = B_\perp^2/k$.

\par
\begin{figure}[hbt]
\begin{center}
\includegraphics*[width=3in]{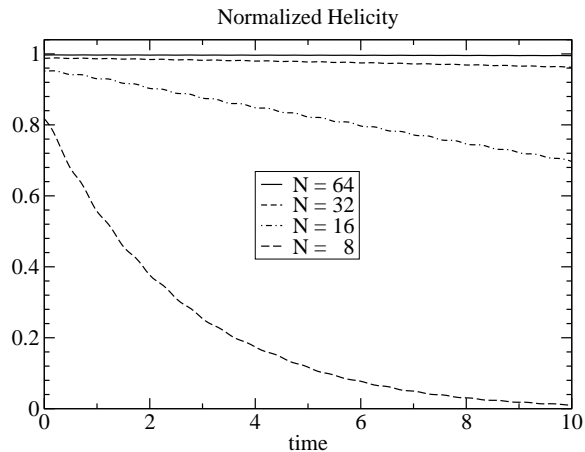}
\end{center}
\caption{Plot of the normalized magnetic helicity 
$\tilde{H} = (k/B_\perp^2) <{\bf b} \cdot {\bf a}>$ as a function of time for
different resolutions.}
\label{fig:cpaw_helicity}
\end{figure}

In figure \ref{fig:cpaw_helicity} we present the time evolution of the
normalized magnetic helicity $\tilde{H} = (k/B_\perp^2) <{\bf b} \cdot
{\bf a}>$ for a traveling Alfv\'en wave using the 6-solve integration
algorithm for a variety of resolutions.  The plots in this figure show
two basic phenomena, dissipation and weak oscillations.  The oscillations
are an indication that the circularly polarized Alfv\'en wave is not
resolved exactly.  As such, certain features regarding the
oscillations are worth mentioning.  First, the oscillation period
$\tau = 1/2$ independent of the grid resolution and whether the
Alfv\'en wave is standing or traveling with respect to the grid.
Second, the amplitude of the oscillations in the helicity varies with
resolution proportional to $N^{-2}$.  Third, the oscillations are
consistent in both amplitude and phase with the independently measured
volume average quantity {}\mbox{${\bf B_0} \cdot <{\bf v}\times{\bf
b}>$}.  These details support the conclusion that the oscillations are
a result of truncation error in resolving the circularly polarized
Alfv\'en wave.

%

\subsection{MHD Riemann Problem Inclined to the Grid}

\par
In this section we present results for the solution of an MHD Riemann
problem in a three dimensional domain.  The Riemann problem is a
favorite test problem for computational algorithms since it can be
chosen to study smooth flows, discontinuous flows, or a combination
thereof.  Moreover, the solution to this problem can, at least in
principle, be calculated exactly allowing one to verify the algorithm
in some parameter regime.  For multidimensional algorithms this can
also be an interesting test problem when the Riemann problem interface
normal direction is chosen such that it has no special orientation
with respect to the computational grid.  In this configuration it
provides a measure of the ability of the computational algorithm to
faithfully reproduce the one-dimensional solution on the large scale,
despite the fact that on the scale of grid cells the flow contains
multidimensional, interacting waves.  In what follows we describe the
initial conditions, translation symmetry and boundary conditions for
this problem and present the solution in a three-dimensional domain
using the 6-solve CTU algorithm.

\par
We begin by choosing a coordinate system $(x_1, x_2, x_3)$ with the
Riemann problem interface located at $x_1=0$ and will use the terms
left and right states to refer to the regions $x_1<0$ and $x_1>0$
respectively.  To map the initial conditions to the computational
domain, we apply the coordinate transformation in equation
{}\ref{eq:wave_coord_trans} with the choice of rotation angles
described below.  This coordinate transformation can be inverted to
read
\begin{eqnarray}
x_1 &=&  x \cos\alpha \cos\beta + y \cos\alpha \sin\beta + z\sin\alpha
\nonumber \\
x_2 &=& -x \sin\beta            + y \cos\beta
\nonumber \\
x_3 &=& -x \sin\alpha \cos\beta - y \sin\alpha \sin\beta + z\cos\alpha ~.
\label{eq:RP_inv_coord_trans}
\end{eqnarray}
Using the fact that the initial conditions and solution to the Riemann
problem are a function of the $x_1$-coordinate alone, the solution
vector $q({\bf x + s}) = q({\bf x})$ for a translation vector ${\bf
s}$ which satisfies $x_1({\bf x + s}) = x_1({\bf x})$.  Making use of
equation (\ref{eq:RP_inv_coord_trans}) we find that the continuous set
of translation vectors ${\bf s}$, for which the solution is invariant,
satisfies the equation
\beq
s_x \cos\alpha \cos\beta + s_y \cos\alpha \sin\beta + s_z \sin\alpha = 0 ~.
\eeq
Now, for the problem at hand we are interested in the discrete set of
translation vectors for which $(s_x, s_y, s_z) = (n_x \delta x, n_y
\delta y, n_z \delta z)$ where $(n_x, n_y, n_z)$ are integers and
$(\delta x, \delta y, \delta z)$ are the grid cell size in each
direction.  Making this substitution, and rearranging terms we find
\beq
n_x + 
n_y \frac{\delta y}{\delta x} \tan\beta + 
n_z \frac{\delta z \tan\alpha}{\delta x \cos\beta} = 0 ~.
\eeq
We next choose the rotation angles $(\alpha, \beta)$ such that 
\beq
\frac{\delta y}{\delta x} \tan\beta = \frac{r_x}{r_y}
%
~~ \textrm{and} ~~ 
%
\frac{\delta z \tan\alpha}{\delta x \cos\beta} = \frac{r_x}{r_z}
\eeq
where $(r_x, r_y, r_z)$ are integers.  With this choice, our equation
for translation invariance becomes
\beq
\frac{n_x}{r_x} + \frac{n_y}{r_y} + \frac{n_z}{r_z} = 0 ~.
\label{eq:3dRP_translate}
\eeq
This is the key relation describing the discrete translation
invariance of the initial conditions, and solution.

\par
There are a couple of interesting implications of this equation which
are of practical importance for this Riemann problem.  First, note
that the translation invariance described by equation
(\ref{eq:3dRP_translate}) was constructed by considering a point
translation symmetry and as such applies equally well to volume and
interface averaged quantities.  That is, there are no approximations
involved in the statement that $q_{i,j,k} = q_{i+n_x, j+n_y, k+n_z}$
for $(n_x, n_y, n_z)$ which satisfy (\ref{eq:3dRP_translate}).
Second, note that one coordinate direction, say the $x$-direction, can
be isolated as the principle simulation direction and the transverse
directions can be made as small as $(r_y, r_z)$.  Finally, note that
the translation invariance relation (\ref{eq:3dRP_translate}) is the
key relation for mapping computational grid cells to ghost cells for
imposing boundary conditions.

\begin{figure}[bht]
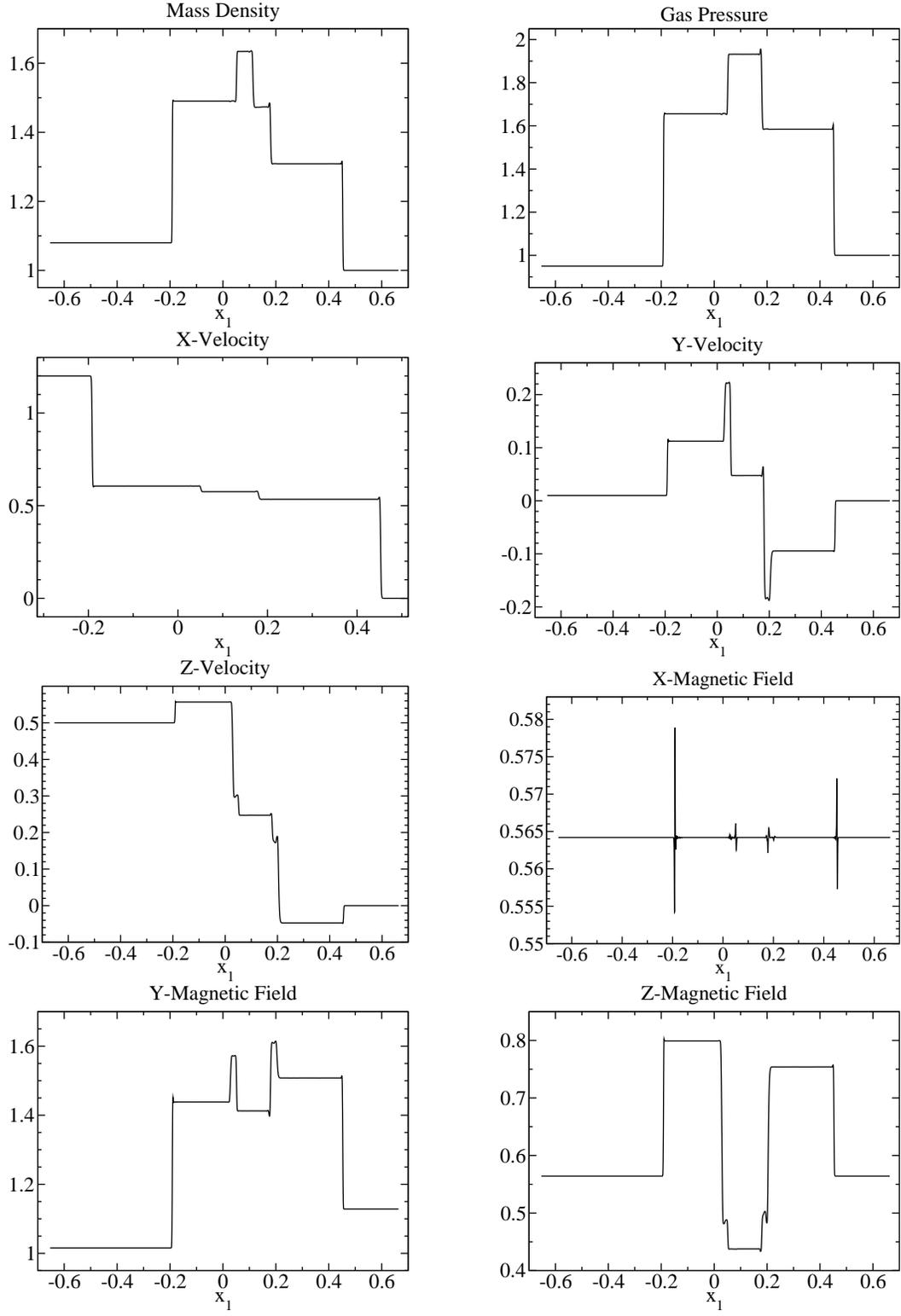

\begin{center}
\includegraphics*[width=0.45\textwidth]{Mass_Density_RJ2a.eps} \hfill
\includegraphics*[width=0.45\textwidth]{Gas_Pressure_RJ2a.eps} \\
\includegraphics*[width=0.45\textwidth]{x_velocity_RJ2a.eps} \hfill
\includegraphics*[width=0.45\textwidth]{y_velocity_RJ2a.eps} \\
\includegraphics*[width=0.45\textwidth]{z_velocity_RJ2a.eps} \hfill
\includegraphics*[width=0.45\textwidth]{x_field_RJ2a.eps} \\
\includegraphics*[width=0.45\textwidth]{y_field_RJ2a.eps} \hfill
\includegraphics*[width=0.45\textwidth]{z_field_RJ2a.eps}
\end{center}
\caption{Solution to the Riemann problem in a direction oblique to the grid.}
\label{fig:3D_RJ2a}
\end{figure}

\par
The specific Riemann problem we consider in this section is presented
in \cite{Ryu-Jones-95} in test problem 2a.  In the $(x_1,~x_2,~x_3)$
coordinate system, the left state is initialized with $\rho=1.08$,
$(v_1,~v_2,~v_3)=(1.2,~0.01,~0.5)$,
$(B_1,~B_2,~B_3)=(2/\sqrt{4\pi},~3.6/\sqrt{4\pi},~2/\sqrt{4\pi})$ and
$P=0.95$.  The right state is initialized with $\rho=1.0$,
$(v_1,~v_2,~v_3)=(0,~0,~0)$,
$(B_1,~B_2,~B_3)=(2/\sqrt{4\pi},~4/\sqrt{4\pi},~2/\sqrt{4\pi})$ and
$P=0.95$.  This problem is then mapped to the 3D domain with the
rotation parameters $(r_x,~r_y,~r_z)=(1,~2,~4)$.  The computational
grid has $(Nx,~Ny,~Nz)=(768,~8,~8)$ grid cells covering the domain
$-0.75 \leq x \leq 0.75$, $0\leq y \leq 1/64$, $0 \leq z \leq 1/64$
and hence has a resolution of $\delta x =\delta y =\delta z =1/512$.

\par
The solution to this Riemann problem at time $=0.2$ is presented in
figure \ref{fig:3D_RJ2a} using the 6-solve CTU algorithm.  These plots
include the cell-center data from every grid cell using the coordinate
transformation in equation (\ref{eq:wave_coord_trans}).  The first
thing to note in these plots is that since $N_y > r_y$ and $N_z > r_z$
there are multiple grid cells with the same cell-center
$x_1$-position.  Therefore, the lack of scatter in these plots
indicates that the algorithm retains the planar symmetry throughout
the simulation.  A comparison of the results presented here to the 1D
solution using the underlying PPM algorithm, with the same resolution,
i.e. $\delta x = 1/512$, indicates that the 3D solution has
dissipation characteristics which are nearly identical to the 1D
algorithm.  The dominant difference between the 1D and 3D solutions is
the presence of oscillations at the slow, Alfv\'en and fast mode
discontinuities.

\par
One question which has received a good deal of attention with this
class of problem is the ability of the computational algorithm to
maintain the parallel component of the magnetic field, $B_1$, equal to
a constant.  We wish to point out here that oscillations are likely
unavoidable unless the orientation of the Riemann problem is chosen to
be aligned in a special direction with respect to the grid.  As
evidence of this fact, we note that the in the initial conditions, the
cell-center $B_1$-component of the magnetic field shows an oscillation
with an amplitude of approximately $8.26\times10^{-3}$ despite the
fact that the interface averaged magnetic fields were initialized with
an ``exact'' integral average using a magnetic vector potential.  This
oscillation is therefore a result of the discretization relating the
cell-center and interface averaged magnetic field components.  In the
initial conditions, as well as the solution at time $=0.2$, the
oscillations in $B_1$ occur wherever the transverse components of the
magnetic field rotate over a small scale such as the initial
discontinuity, and the resultant fast, Alfv\'en and slow mode
discontinuities.  Finally, we note that just as in the 2D paper
\cite{GS05}, the oscillations in the parallel component of the
magnetic field can be eliminated by restricting the solution to
``macrocells''.  This operation effectively aligns the the
$x_1$-direction with the macrocell $[1,1,1]$ direction.

\subsection{MHD Blast Wave}

Another problem which has been computed by a number of authors is the
explosion of a centrally over pressurized region into a low pressure,
low $\beta$ ambient medium.  This is an interesting problem in the
sense that it combines shocked flows, smooth flow regions, and strong
magnetic fields.  While the results of this test are not particularly
quantitative in their measure of the accuracy, this test is a good
measure of the robustness of the integration algorithm.  Variants on
this problem have been presented by a number of authors
{}\cite{Zachary94,Balsara-Spicer,Londrillo-DelZanna00,GS05} and here
we choose to use the parameters given by \cite{Londrillo-DelZanna00}
for a three-dimensional domain.

\par
The computational domain extends from $-0.5 \le x \le 0.5$, $-0.5 \le
y \le 0.5$ and $-0.5 \le z \le 0.5$.  The density $\rho=1$, the
velocity ${\bf v=0}$, and the magnetic field components $B_x = B_z =
10/\sqrt{2}$ and $B_y=0$.  Within a sphere of radius $R = 0.125$ about
the origin the gas pressure $P=100$ and $\beta = 2 P/B^2 = 2$.
Outside of this sphere, the gas pressure $P=1$ and $\beta = 2\times
10^{-2}$.  These initial conditions are evolved until a time $t=0.02$
using a $200^3$ computational grid.

\par
\begin{figure}[hbt]
\begin{center}
\includegraphics*[width=2.5 in]{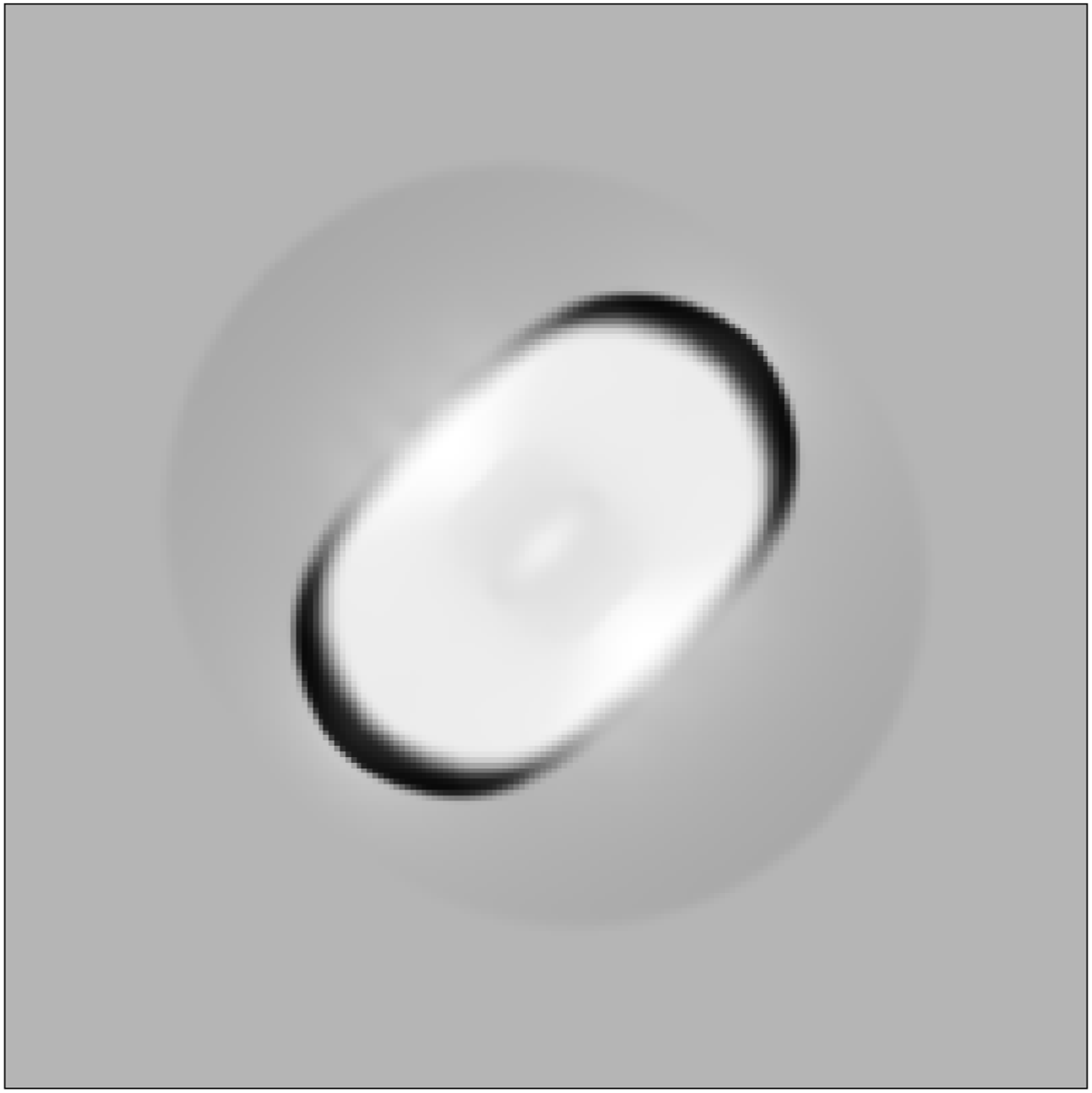} \hfill
\includegraphics*[width=2.5 in]{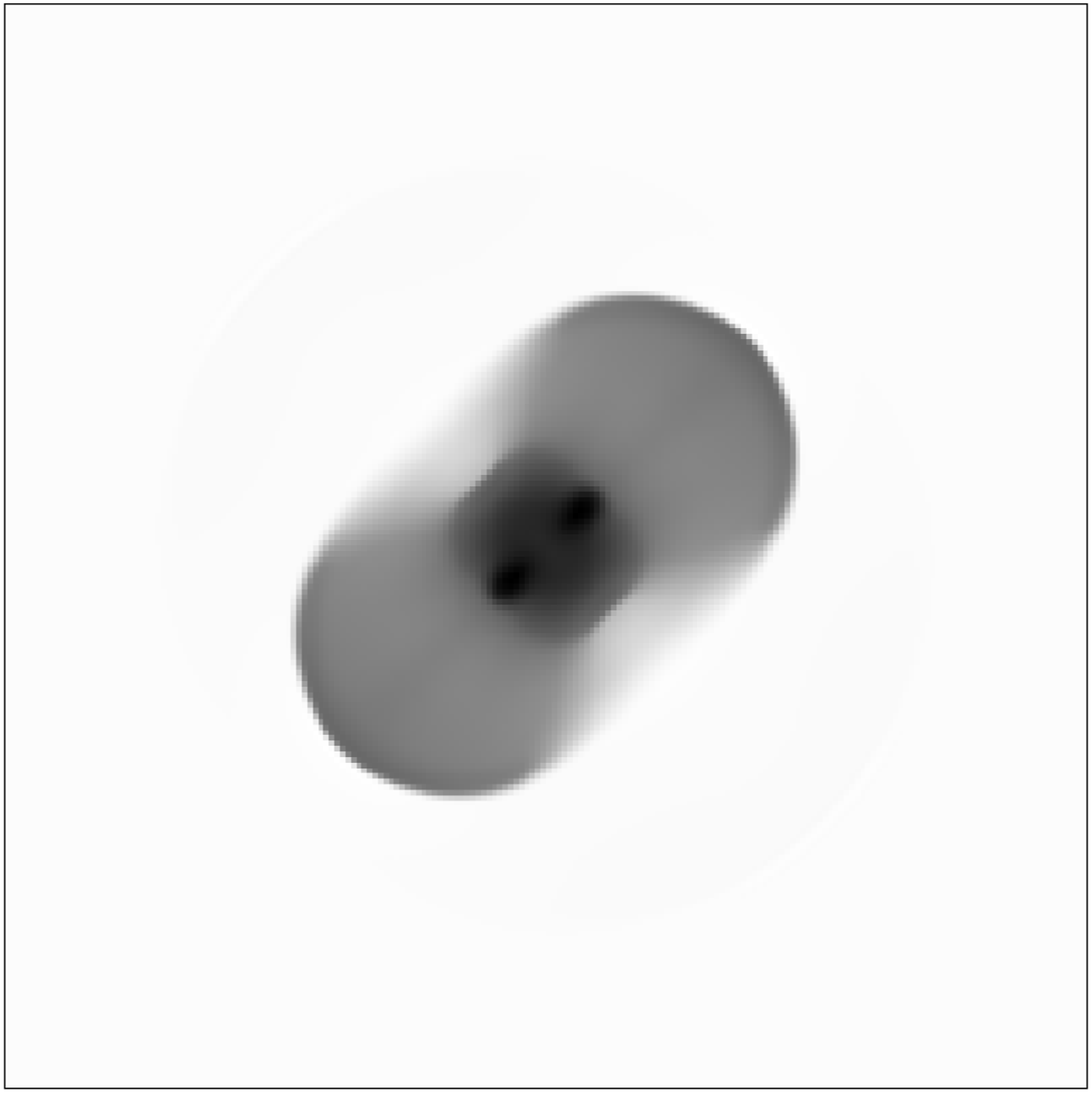} \\[0.25in]
\includegraphics*[width=2.5 in]{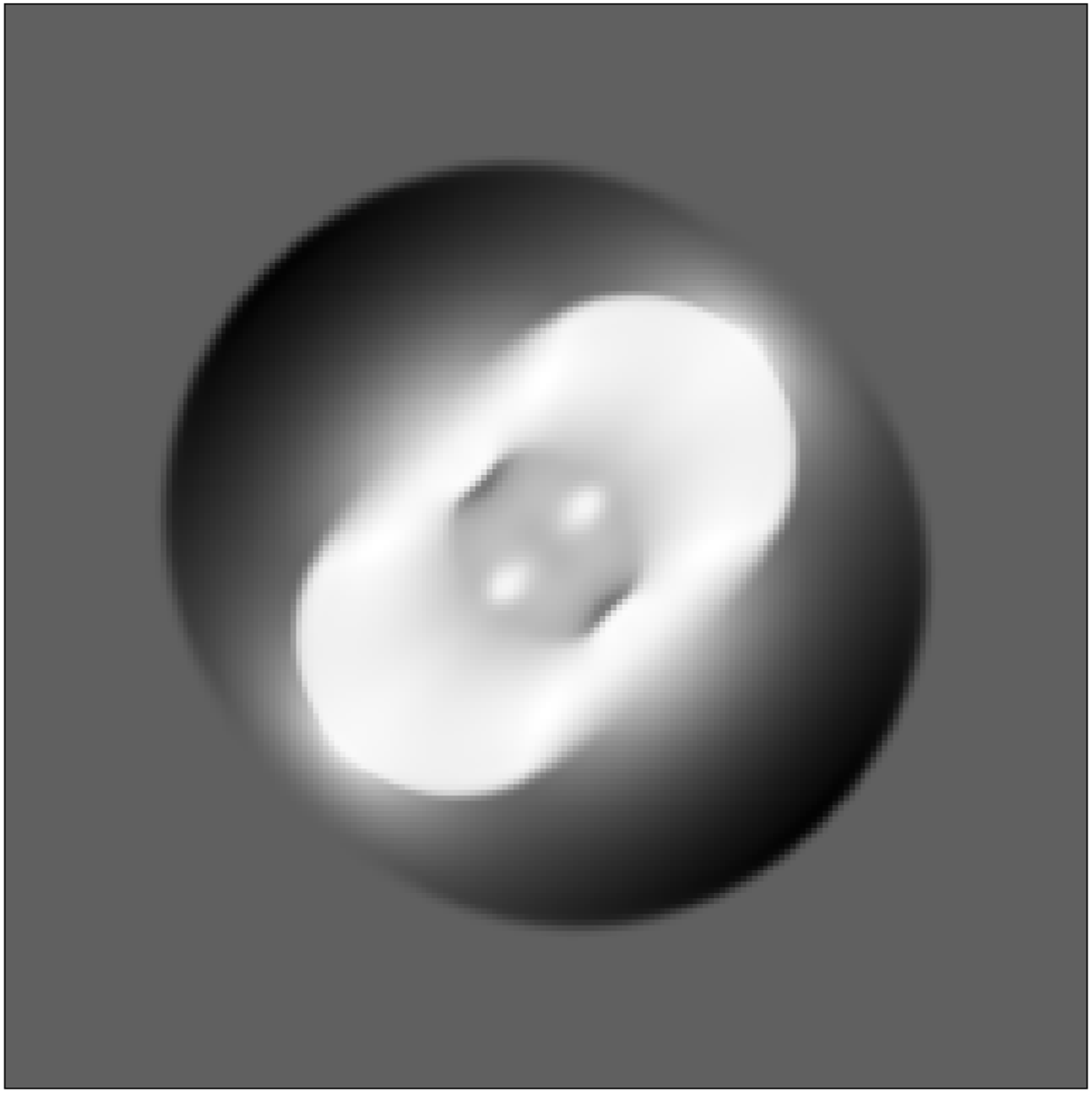} \hfill
\includegraphics*[width=2.5 in]{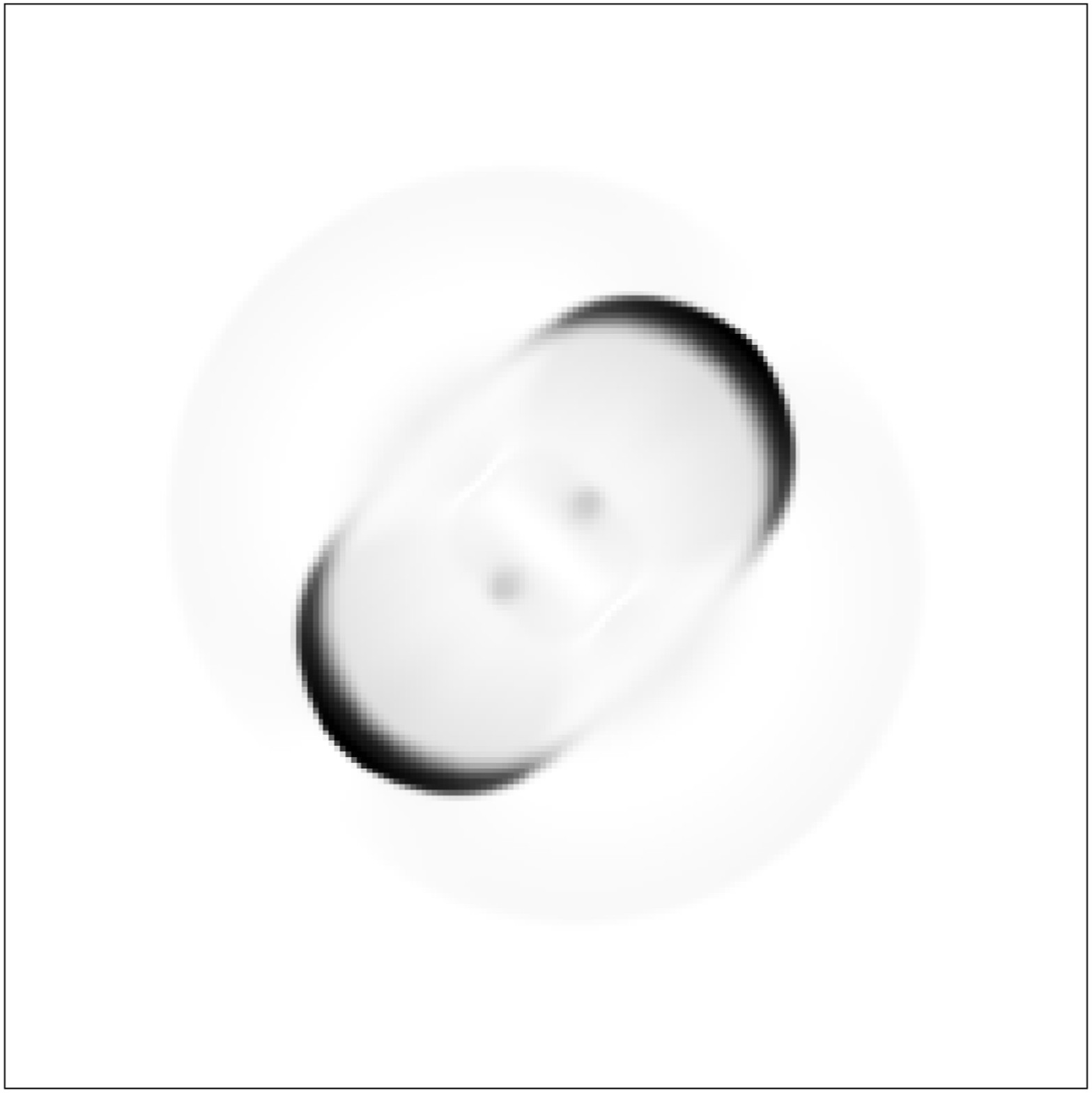} \\
\end{center}
\caption{Linearly scaled grey-scale images of the evolved state 
(time=0.02) for the MHD blast wave problem.  The density (\emph{top
left}) ranges from 0.190 (white) - 2.98 (black).  The gas pressure
(\emph{top right}) ranges from 1.0 (white) - 42.4 (black). The
magnetic energy density (\emph{bottom left}) ranges from 25.2 (white)
- 64.9 (black).  The kinetic energy density (\emph{bottom right})
ranges from 0.0 (white) - 33.1 (black).}
\label{fig:mhd_blast}
\end{figure}

In figure \ref{fig:mhd_blast} we present images of the density,
pressure, magnetic and kinetic energy density sliced along the $y=0$
plane at the end time.  The general structure of the solution is the
same as one finds in the 2D calculation.  Namely, the outermost
surface in this expanding shell is a fast-shock which is only weakly
compressive and energetically is dominated by the magnetic field.
Interior to this, one finds two dense shells of gas which propagate
parallel to the magnetic field.  These shells are bounded by a
slow-mode shock and contact surface (separating the initially hot,
interior gas from the surrounding cool ambient medium) on the outer
and inner surfaces respectively.  The maximum compression of the
ambient gas by the slow-mode shock is approximately 3.3, the same as
was found in the 2D calculation.  The fact that the 2D and 3D
calculations show quantitatively similar compression in the show-mode
shock is an indication that their motion is approximately
one-dimensional, i.e. parallel to the magnetic field.

\par
The results of this section are interesting from the point of view
that they demonstrate that the 6-solve integration algorithm is a
robust algorithm, capable of evolving shocked flows with
$\beta\sim10^{-2}$.  Moreover, since the integration algorithm is
unsplit, it preserves the symmetry of the initial conditions
naturally.

%

\section{Conclusion}
\label{sec:Conclusions}

In this paper we have presented a three-dimensional MHD integration
algorithm which combines the (6-solve) Corner Transport Upwind
integration algorithm with the method of Constrained Transport for
evolving the magnetic field.  This integration algorithm is a natural
extension, and generalization of the two-dimensional algorithm
{}\cite{GS05}.  In addition we have outlined the essential elements to
constructing a 12-solve CTU with CT integration algorithm for MHD and
included results of this algorithm in \S\ref{sec:tests}.  Both the
6-solve and 12-solve algorithms are found to be accurate and robust
for approximately the same computational cost.  As a result, we
generally prefer the 6-solve algorithm as a result of its simplicity
and smaller memory footprint.

\par
The three-dimensional MHD PPM interface states algorithm presented in
this paper is a new and essential element of the integration
algorithms.  We have shown here that this is a natural extension of
the 2D MHD PPM interface states algorithm presented in {}\cite{GS05}
and that it reduces identically to the 2D algorithm in the
grid-aligned, plane-parallel limit.  The 3D MHD PPM interface states
algorithm was designed in such a way as to satisfy a multidimensional
balance law involving what we have referred to here as MHD source
terms.  Failure to satisfy this balance law is found to result in
erroneous and secular evolution of the magnetic field under quite
general conditions, e.g. the advection of a high $\beta$ magnetic
field loop.  

\par
We have also presented a variety of test results for both the 6- and
12-solve MHD CTU CT integration algorithms.  These test problems were
selected so as to enable a comparison with previously published
results, as well as to introduce new, quantitative measures of the the
solution accuracy.  One interesting result of these tests is the
observation that the dominant difference in the L1 error for the 6-
and 12-solve algorithm convergence on smooth wave propagation is
attributable to the CFL number dependence.  Throughout this section we
have included the necessary information so as to enable other
researchers involved in developing or applying MHD algorithms to make
a quantitative, as well as qualitative, comparison with the results in
this paper.

\par
Finally, it is worth noting that the integration algorithms presented
here have been thoroughly tested on a great many test problems not
included here.  These include problems which are also of interest for
their scientific merit.  Examples include a study of the
magneto-rotational instability \cite{BH98RMP} and the MHD Raleigh
Taylor instability \cite{SG06}.  In a future paper we will detail our
approach to combining the integration algorithms presented here with
the methods of static and adaptive mesh refinement.

\section{Acknowledgments}

Simulations were performed on the Sun Grid computational facility,
Teragrid cluster at NCSA, the IBM Blue Gene at Princeton University,
and on computational facilities supported by NSF grant
AST-0216105. Financial support from DoE grant DE-FG52-06NA26217 is
acknowledged.

\appendix

\section{Linear Wave Right Eigenvectors}

In order to enable others to perform the linear wave convergence test
presented in section \ref{sec:linear_wave} and compare their results
in a quantitative manner, we include the numerical values for the
right eigenvectors here.  In the wave-aligned coordinate system
$(x_1,~x_2,~x_3)$ the conserved variable vector and right eigenvectors
(labeled according to their propagation velocity) are given by
\bdm
q = \left (
\begin{array}{c}
\rho \\
\rho v_1 \\
\rho v_2 \\
\rho v_3 \\
B_1 \\
B_2 \\
B_3 \\
E
\end{array}
\right ),~~~
%
%
R_{\pm c_f} = \frac{1}{2\sqrt{5}} \left (
\begin{array}{c}
2 \\
\pm 4 \\
\mp 2 \\
0 \\
0 \\
4 \\
0 \\
9
\end{array}
\right ),~~~
%
%
R_{\pm c_a} = \left (
\begin{array}{c}
0 \\
0 \\
0 \\
\mp 1 \\
0 \\
0 \\ 
1 \\ 
0 \\
\end{array}
\right ),
\edm
\beq
R_{\pm c_s} = \frac{1}{2\sqrt{5}} \left (
\begin{array}{c}
4 \\
\pm 2 \\
\pm 4 \\
0 \\
0 \\
-2 \\
0 \\
3
\end{array}
\right ),~~~
%
%
R_{v_1} = \frac{1}{2} \left (
\begin{array}{c}
2 \\
2 \\
0 \\
0 \\
0 \\
0 \\
0 \\
1
\end{array}
\right ).
\eeq
%


\end{document}